\documentstyle[12pt]{article}
\newcommand{\slq}{\raise.15ex\hbox{$/$}\kern-.57em\hbox{$q$}}
\newcommand{\slp}{\raise.15ex\hbox{$/$}\kern-.57em\hbox{$p$}}
\newcommand{\be}{\begin{equation}}
\newcommand{\ee}{\end{equation}}
\newcommand{\bear}{\begin{eqnarray}}
\newcommand{\ear}{\end{eqnarray}}

\newcommand{\gf}{{\rm{gf}}}
\newcommand{\gh}{{\rm{gh}}}

\newcommand{\Tr}{{\rm{Tr}}}

\newcommand{\R}{{\cal{R}}}
\date{}
\renewcommand{\theequation}{\arabic{section}.\arabic{equation}}

\begin{document}
\begin{titlepage}
\begin{flushright}
HD--THEP--95--2
\end{flushright}
\quad\\
\vspace{1.8cm}
\begin{center}
{\bf\LARGE Effective Quark Interactions}\\
\medskip
{\bf\LARGE from QCD}\\
\vspace{1cm}
Christof Wetterich\\
\bigskip
Institut  f\"ur Theoretische Physik\\
Universit\"at Heidelberg\\
Philosophenweg 16, D-69120 Heidelberg\\
\vspace{1cm}
{\bf Abstract}
\end{center}
We propose a new method for an analytical, non-perturbative
computation of effective quark interactions from QCD. It is
based on an exact flow equation which describes the scale
dependence of the effective average action for quarks in presence
of gluons.

\end{titlepage}
\newpage
\section{Introduction}
Quantum chromodynamics as the theory of strong interactions shows very
different facets at short and long distances. The high momentum
behaviour is governed by asymptotic freedom \cite{1E} and the
relevant degrees of freedom are quarks and gluons. In contrast, the
particles which are observed at large length scales (larger than
1 fm) are mesons and hadrons. The interactions of the pseudoscalar mesons
are modeled by chiral perturbation theory \cite{2E}. The
corresponding nonlinear $\sigma$-model shares the flavour symmetries of
perturbative QCD. It is believed that the free phenomenological parameters
of this model can ultimately be computed from the action of QCD, but this
is not a simple task.

Recently, the transition from quark degrees of freedom to meson
degrees of freedom has been described by a nonperturbative flow equation
\cite{Ell}. It is based on the concept of the effective average action
$\Gamma_k$ \cite{W1} for which only quantum fluctuations with
(covariant) momenta $q^2>k^2$ are integrated out. The average action
is the effective action for averages of fields. It acts like a
microscope by which we can look at the theory at different length
scales, with a ``resolution'' given by the scale $k^{-1}$. For
$k=0$ one recovers the usual effective action, i.e. the generating
functional of the 1PI Green functions, whereas for $k\to\infty$
$\Gamma_k$
equals the classical action.
The dependence of $\Gamma_k$
on the scale $k$ is described by an exact nonperturbative flow
equation \cite{4E}. By the introduction of composite fields
for the mesons the original equation formulated for quarks can be
transmuted into an equivalent exact flow equation involving also
the mesons \cite{Ell}.

In a first attempt the chiral condensate $<\bar\psi\psi>$ and the
pion decay
constant $f_\pi$ were computed along these lines \cite{Ell}. In this
approach the gluons have not been considered explicitly. Their effect
was encoded in a phenomenologically motivated four-quark interaction,
which
may be thought as the result of integrating out the gluons in
the defining functional integral. If one aims at a computation
of the effective parameters of chiral perturbation theory from the
QCD action, this shortcut has to be removed. A full computation should
start at short distances with the quark-gluon description of QCD and
systematically account for the quantum fluctuations of the gluon field.

Conceptually, one may integrate out the gluons at once and end with
an effective quark theory. In practice, this seems almost impossible:
A determination of the resulting complicated nonlocal quark
interactions
requires more or less a complete solution of QCD. On the other hand,
the form of the flow equation at a given scale $k$ is not sensitive
to details of physics at momentum scales much below $k$. Only the
effective propagators and vertices for modes with momenta $q^2\approx
k^2$ play a role for this equation. A typical scale where the light
mesons
form is around 700 MeV  and therefore substantially higher than the
confinement scale. This situation further improves for problems
involving heavier quarks as for example the charmed or beauty mesons.
One may hope that very detailed features of confinement are not
needed for an understanding of the mesons. An appropriate tool would
therefore be a method which only integrates out the gluon fluctuations
with momenta $q^2>k^2$ instead of addressing
the much more complicated problem
of integrating out all gluon fluctuations at once. Formally, it is
easy to do this at a given scale $k_1$. One needs to compute the
effective
action for quarks and gluons $\Gamma_{k_1}[\psi,A]$ at this scale.
Solving the
classical field equation for the gluon field $A$ in dependence on
$\psi$ and
inserting this classical solution into $\Gamma_{k1}[\psi,A]$ yields
exactly an effective average action
$\Gamma_{k_1}[\psi]$ involving
only the quark fields. The quark-effective action $\Gamma_{k_1}[\psi]$
may then be used as an initial value for solving a pure fermionic flow
equation for $k<k_1$. Obviously, the shortcoming of such an approach
ist the complete omission of the effects of gluon fluctuations with
$q^2<k_1^2$. Choosing a different scale $k_2$ for eliminating the gluons
will lead to a different result. Such a sharp transition between the
quark-gluon system and a description involving only quarks necessarily
introduces a certain degree of arbritariness.

We propose here a more refined method which changes the classical
solution for the gluon field in the course of the evolution towards
lower $k$, thus reflecting the change in the form of $\Gamma_k[\psi,A]$.
The result is a smooth procedure for integrating out the gluons,
where at every scale $k$ all gluon fluctuations with $q^2>k^2$ are
included. Nevertheless, we obtain a flow equation for the quark
effective
average action $\Gamma_k[\psi]$, where gluon fields do not appear
explicitly. As a consequence of the inclusion of contributions from
additional gluon fluctuations as the scale is lowered, correction
terms appear in the flow equation for the quark interactions. In
particular, we choose here a formulation where the $k$-dependent
classical solution for $A$ as a functional of $\psi$ includes an
effective infrared cutoff $\sim k$. As a consequence, the only
nonlocalities
in $\Gamma_k[\psi]$ concern length scales shorter than $k^{-1}$. For
the fermionic low momentum modes $\Gamma_k[\psi]$ is an
effectively local action. For example, a derivative expansion is
meaningful for $q^2\ll k^2$. The expected nonlocalities arising from
the complete elimination of gluon fields (for example a four-quark
interaction $\sim\frac{1}{q^2}$ appearing already in the Born
approximation) build up only step by step as $k$ is lowered to
zero. As a result of this method we will end with an exact
nonperturbative
flow equation for the scale dependence of $\Gamma_k[\psi]$.
Approximations will be needed to solve this equation but they are
not limited to perturbative concepts. The correction terms reflecting the
gluon fluctuations require limited knowledge about the effective gluon
propagator and vertices. It is hoped that rather crude approximations
for the gluonic vertices can already lead to satisfactory results.

In sect. 2 we first demonstrate our formalism for a simple model of two
scalar fields. The flow of the effective average action for one of
the scalar fields obtains by integrating out the other scalar field
at any scale $k$. Subsequently this
is generalized to quarks and gluons. We also
give a first demonstration how this formalism describes
the flow of the two- and four-point function in the effective quark     
theory. The special case of heavy quarks is addressed in sect. 3. Here
we argue that the evolution of the gauge field propagator is needed
in this limit. In sect. 4 we compute the corresponding flow equation
and discuss the scale dependence of the gluon propagator.
In sect. 5 we collect all the ingredients needed
for the flow equation in case of light quarks. Finally, our conclusions
are contained in sect. 6.

\section{Reduction of degrees of freedom}
\setcounter{equation}{0}
First we consider for simplicity two types of scalar fields,
$\varphi$ and $\psi$. We want to develop a formalism how to translate
evolution equations for the effective average action for $\varphi$
and $\psi$
into corresponding equations involving only $\psi$. The reader may
associate $\varphi$ with the gluon fields and $\psi$ with the quark
fields.
Our aim is then the construction of the effective average action for
quarks
out of the coupled quark-gluon system. This amounts to integrating
out the
gluonic degrees of freedom represented in the simplified model
by $\varphi$. We start with the scale-dependent generating
functional
for the connected Green functions
\be\label{2.1}
W_k[J,K]=\ln\int D\varphi'D\psi'\exp-
\bigl\lbrace S[\varphi',\psi']
+\Delta^{(\varphi)}_kS[\varphi']+\Delta_k^{(\psi)}S[\psi']-J^\dagger
\varphi'-K^\dagger\psi'\bigr\rbrace\ee
Here we denote the degrees of freedom contained in $\varphi'$
(for example
the Fourier modes) by ${\varphi'}^\alpha$
and similar for $\psi'$, $J$ and $K$, with\footnote{We use indices
$\alpha,\alpha'$ etc. for $\varphi$ and $\beta,\beta'$ etc. for $\psi$.}
$J^\dagger\varphi'=J^*_\alpha{\varphi'}^\alpha,\quad K^\dagger\psi'=
K_\beta^*{\psi'}^\beta$.
We have introduced an infrared cutoff quadratic in the fields
\be\label{2.3}
\Delta^{(\varphi)}_kS[\varphi']=\frac{1}{2}{\varphi'}^\dagger
R^{(\varphi)}_k\varphi'\ee
and similar for $\psi$. This suppresses the contribution of fluctuations
with small momenta $q^2<k^2$ to the functional integral (\ref{2.1}).
Typically
$R^{(\varphi)}_k,R_k^{(\psi)}$ are functions of $q^2$ as, for example,
\be\label{2.5}
R_k^{(\varphi)}=\frac{Z_kq^2\exp\left(-\frac{q^2}{k^2}\right)}{1-\exp
\left(-\frac{q^2}{k^2}\right)}\ee
which acts like a mass term $R_k^{(\varphi)}\sim Z_kk^2$
for $q^2\ll k^2$.
The effective average action $\Gamma_k[\varphi,\psi]$ is related to
the Legendre transform of $W_k[J,K]$
\be\label{2.6}
\tilde\Gamma_k[\varphi,\psi]=-W_k[J,K]+J^\dagger\varphi+K^\dagger\psi\ee
by subtracting the infrared cutoff term
\be\label{2.8}
\Gamma_k[\varphi,\psi]=\tilde\Gamma_k[\varphi,\psi]-\Delta_k^{(\varphi)}S
[\varphi]-\Delta_k^{(\psi)}S[\psi]\ee
For $k\to0$ the infrared cutoff $\Delta_kS=\Delta_k^{(\varphi)}S+\Delta_k
^{(\psi)}S$ vanishes and $\Gamma_0$ is the usual generating function for
the 1PI Green functions. Using the quadratic form of $\Delta_kS$ it is
straightforward to derive an exact non-perturbative
evolution equation for the dependence
of the effective average action on the scale $k$ $(t=\ln k)$ \cite{4E}
\be\label{2.9}
\frac{\partial\Gamma_k}{\partial t}=\frac{1}{2}Tr\left\lbrace
(\tilde\Gamma_k^{(2)})^{-1}\frac{\partial R_k}{\partial t}
\right\rbrace\ee
Here $\tilde\Gamma_k^{(2)}=\Gamma_k^{(2)}+R_k$ and the inverse propagator
$\Gamma_k^{(2)}$ is the second functional derivative of $\Gamma_k$ with
respect to the fields. The matrix
$R_k=R_k^{(\varphi)}+R_k^{(\psi)}$ is block diagonal in $\varphi$ and
$\psi$
spaces.
The presence of the infrared cutoff $R_k$ in
$\tilde \Gamma_k^{(2)}$ guarantees infrared finiteness for the momentum
integral implied by the trace even in case of massless modes. Ultraviolet
finiteness is guaranteed by the exponential decay of $\partial R_k/\partial
t$ (\ref{2.5}). A solution of the flow equation (\ref{2.9}) interpolates
between the classical action for $k\to\infty$ (or $k$ equal to
some ultraviolet cutoff $\Lambda$) and the effective action for $k\to0$.

The generating functional for the connected Green functions for $\psi$
obtains
>from (\ref{2.1}) for $J=0$
\be\label{2.10}
W_k[K]\equiv W_k[J=0,K]\ee
Correspondingly, we may
introduce an effective action expressed only in terms of
$\psi$
\be\label{2.11}
\tilde\Gamma_k[\psi]=\tilde\Gamma_k[\varphi_k[\psi],\psi]\ee
\bear\label{2.12}
\Gamma_k[\psi]&=&\tilde\Gamma_k[\psi]-\Delta_k^
{(\psi)}S[\psi]\nonumber\\
&=&\Gamma_k[\varphi_k[\psi],\psi]+\Delta_k^{(\varphi)}S[\varphi_k[\psi]]
\ear
by inserting the $k$-dependent solution of the field equation
\be\label{2.13}
\frac{\partial\tilde\Gamma_k[\varphi,\psi]}{\partial\varphi^\alpha}
_{|\varphi_k[\psi]}=0\ee
This defines $\varphi_k$ as a $k$-dependent functional of $\psi$.
It is easy to verify that $\tilde\Gamma_k[\psi]$ is the Legendre
transform
of $W_k[K]$ (\ref{2.10}). One concludes for $k\to0$ that $\Gamma_0[\psi]$
is the generating functional for the 1PI Green functions for $\psi$.

We want to employ the flow equation (\ref{2.9}) for finding the
$k$-dependence
of $\Gamma_k[\psi]$. In addition to the corresponding equation for only
one type of fields we have here additional contributions from the
$k$-dependence
of $\Delta_k^{(\varphi)}S$ in (\ref{2.1}).
The evolution equation for
$\Gamma_k[\psi]$ can now be obtained by performing in eq. (2.9)
a variable transformation which amounts to a shift of $\varphi$
around $\varphi_k[\psi],
\hat\varphi^\alpha=\varphi^\alpha-
\varphi^\alpha_k[\psi]$. One obtains
\bear\label{2.17}
&&\frac{\partial}{\partial t}\Gamma_k[\psi]=\frac{1}{2}\left(\Gamma_k^{(2)}
[\psi]+R_k^{(\psi)}\right)_{\ \ \ \beta'}^{-1\beta}\left\lbrace
\left(\frac{\partial
R_k^{(\psi)}}{\partial t}\right)^{\beta'}_{\ \beta}+
\frac{\partial\varphi^*_{k\alpha'}}{\partial\psi^*_{\beta'}}
\left(\frac{\partial
R_k^{(\varphi)}}{\partial t}\right)^{\alpha'}_{\ \alpha}
\frac{\partial\varphi_k^\alpha}{\partial\psi^\beta}\right\rbrace\nonumber\\
&&+\frac{1}{2}\varphi^*_{k\alpha'}[\psi]\left(\frac{\partial
R_k^{(\varphi)}}
{\partial t}\right)^{\alpha'}_{\ \ \alpha}\varphi_k^\alpha[\psi]
+\frac{1}{2}\left(\tilde\Gamma^{(2)}_k[\varphi=\varphi_k,\psi]\right)
^{-1\alpha}_{\ \ \ \alpha'}\left(\frac{\partial R_k^{(\varphi)}}{\partial t}
\right)^{\alpha'}_{\ \alpha}\ear
It is easy to verify that this equation reduces in the limit
$R_k^{(\varphi)}=0$ to the equivalent of eq. (\ref{2.9}) for fields
$\psi$ only. The corrections in the first two terms involve the explicit
form of the ``classical solution'' $\varphi_k[\psi]$. If one
is interested in
1PI Green functions for $\psi$ with a given number of external legs
one only
needs a polynomial expansion of $\varphi_k[\psi]$ up to a given order.
For example, the evolution of the term $\sim\psi^4$ in $\Gamma_k[\psi]$
needs the classical solution up to the order $\psi^4$ if the series
$\varphi_k[\psi]$ starts with a term quadratic in $\psi$. Additional
knowledge of the form of $\Gamma_k[\hat\varphi,\psi]$ beyond its value
for $\hat\varphi=0$ is only needed for the last correction term
in the form
of
\be\label{2.18}
\left(\hat\Gamma_k^{(2)}[0,\psi]\right)^{\alpha'}_{\ \ \alpha}=
\frac{\partial^2
\Gamma_k[\varphi,\psi]}{\partial\hat\varphi^*_{\alpha'}
\partial\hat\varphi^\alpha}
{}_{|\hat\varphi=0}
+\left(R_k
^{(\varphi)}\right)^{\alpha'}_{\ \ \alpha}\ee
Only the $\psi$-dependence of the effective $\hat\varphi$-propagator plays
a role for the study of 1PI functions for $\psi$.

Let us next apply the general flow equation (\ref{2.17}
explicitly to the quark-gluon system.
If $\psi$ is a Grassmann variable as appropriate for fermions the matrix
$R_k$ in (\ref{2.9}) becomes
\be\label{3.1}
R_k=R_k^{(\varphi)}-R_k^{(\psi)}\ee
Also $\psi^*$ should be replaced by $\bar\psi$ and the index summation over
$\beta$ should involve both $\psi$ and $\bar\psi$ separately. For the gauge  
fields we will
choose here a formulation with explicit ghost variables in close
analogy, but slightly different from the formulation in ref. \cite{Reu}.
This
makes our formulation as close as possible to the language of standard
perturbation theory. Details can be found in the appendix.

We start with the action including a gauge-fixing term in the background
gauge and a corresponding action for the anticommuting ghost fields
$\xi,\bar\xi$
\be\label{3.2}
\hat S[\psi',\xi',a;\bar A]=S[\psi',A']+S_{gf}[a;\bar A]+S_{gh}
[\xi',a;\bar A]\ee
Here $S$ is a gauge invariant functional of the fermion fields $\psi,
\bar\psi$ and the gauge field
\be\label{3.3}
A'_\mu=\bar A_\mu+a_\mu.
\ee
The background gauge field $\bar A_\mu$ appears in the gauge fixing and
ghost terms
\bear
S_{gf}&=&\frac{1}{2\alpha}\int d^d x G^*_z G^z\label{3.4}\\
G^z&=&(D^\mu[\bar A])^z_{\ y}  a^y_\mu\label{3.5}\\
S_{gh}&=&\int d^d x\bar\xi'_y(-D^\mu[\bar A] D_\mu[\bar A+a])^y_{\ z}
\xi^{\prime z}\label{3.6}
\ear
Here $D_\mu[\bar A]$ is the covariant derivative in the adjoint
representation
in presence of  the background gauge field $\bar A$. The generating
functional
for the connected Green functions is defined as usual
\be\label{3.7}
W[\eta,\zeta,K;\bar A]=\int {\cal D}\psi' {\cal D}\bar\psi'{\cal D}\xi'
{\cal D}\bar\xi' {\cal D}a\exp-\{\hat S-\int d^d x[\bar\eta \psi'+  
\eta\bar\psi'+\bar\zeta\xi'+\zeta\bar\xi'+Ka]\}\ee
We have introduced here also sources $\zeta$ for the ghost fields and
note that the source $K^\mu_z$ couples to the gauge field fluctuation
$a^z_\mu$ and therefore transforms homogeneously under gauge
transformations
as an adjoint tensor. The $k$-dependent version $W_k$ obtains from $W$ by
adding to $\hat S$ the infrared cutoff piece
\be\label{3.8}
\Delta_k S=\Delta^{(\psi)}_k S+\Delta ^{(A)}_kS+\Delta^{(gh)}_k S
\ee
Here the fermionic cutoff reads
\bear\label{3.9}
\Delta^{(\psi)}_k S&=&\bar\psi'_{\beta'}
(R^{(\psi)}_k)^{\beta'}
_{\ \beta}\psi^{\prime\beta}\nonumber\\
&=&\int d^d x\bar\psi' Z_{\psi,k}
(i\gamma^\mu D_\mu[\bar A])\ r_k^{(\psi)}(-D^2[\bar A]/k^2)\psi'
\ear
with $D_\mu$ the covariant derivative in the appropriate representation
$(D^2=D_\mu D^\mu)$ and $r^{(\psi)}_k$ a dimensionless function.  For the
gauge field cutoff we choose
\bear\label{3.10}
\Delta^{(A)}_kS&=&\frac{1}{2} a^*_{\alpha'}(R_k)^{\alpha'}_{\ \alpha}
a^\alpha\nonumber\\
&=& \frac{1}{2} \int d^d xa^y_\nu \left[{\cal D}[\bar A]\ r_k^{(A)}
\left(\frac{{\cal Z}^{-1}_{A,k}{\cal D}[\bar A]}{k^2}\right)\right]
^{y\mu}_{\nu z} a^z_\mu
\ear
with ${\cal D}[\bar A]$ an appropriate operator generalizing a covariant
Laplacian in the adjoint representation which will be explained below.
The matrix ${\cal Z}_{A,k}$ accounts for an appropriate wave function
renormalization.  Finally, we take for the ghosts
\bear\label{3.11}
\Delta^{(gh)}_kS&=&\bar\xi'_{\gamma'}(R^{(gh)}_k)^{\gamma'}_{\ \ \gamma}
\xi^{\prime
\gamma}\nonumber\\
&=&\int d^d x\bar\xi'_y [Z_{gh,k} {\cal D}_s[\bar A] r_k^{(gh)}
({\cal D}_s[\bar A]/k^2)]^y_{\  z}\xi^{\prime z}
\ear
with ${\cal D}_s[\bar A]=- D^2[\bar A]$ in the adjoint representation.
A good choice for the dimensionless function $r_k$ is\footnote{The function
$r_k^{(\psi)}$ may be chosen differently from (\ref{3.12}) in order to
avoid that $R_k$ diverges for vanishing covariant momenta.}
\be\label{3.12}
r_k(y)=\frac{e^{-y}}{1-e^{-y}}\ee
such that
\be\label{3.13}
\lim_{{\cal D}\to 0} R_k=Z_k k^2
\ee
The $k$-dependent functions $Z_{\psi,k,}{\cal Z}_{A,k}$ and $Z_{gh,k}$
will be adapted to corresponding wave function renormalization constants
in the kinetic terms for the fermions, gauge fields and ghosts.
 In principle, they can depend on the background field $\bar A$.
The infrared cutoff piece $\Delta_k S$  cuts off all quantum fluctuations
with covariant momenta smaller than $k$ in the functional integral
defining $W_k$. For covariant momenta larger than $k$ the infrared cutoff
is ineffective and its contribution to the propagator is exponentially
suppressed.

Performing a Legendre transform  and subtracting the IR cutoff piece
again (c.f. (\ref{2.6}), (\ref{2.8})) we arrive at the effective average
action $\Gamma_k[\psi,\xi,A,\bar A]$, where $A=\bar A+\bar a$ and
$\bar a$ is
conjugate to $K$. The dependence of $\Gamma_k$ on the scale $k$
is described by an exact evolution equation analogous to eq. (\ref{2.9}),
with a negative sign for the contributions $\sim R_k^{(\psi)}$ and
$R_k^{(gh)}$. It is derived in the appendix (A.12).
We note that  $\Gamma_k$ only involves terms with an even
number of ghost fields due to the symmetry $\bar\xi'\to -\bar\xi',
\xi'\to-\xi'$ of the  $S_{gh}$ and $\Delta_k^{(gh)} S$. In consequence,
the ghost field equations
\be\label{3.14}
\frac{\delta\Gamma_k}{\delta\bar\xi}=0,\ \frac{\delta \Gamma_k}
{\delta\xi}=0
\ee
have always the solution $\bar\xi=\xi=0$. We therefore can extract the
propagators and vertices for the physical particles from the effective
action for $\bar\xi=\xi=0$:
\be\label{3.15}
\Gamma_k[\psi,A,\bar A]=\Gamma_k[\psi,0,A,\bar A]\ee
Nevertheless, the evolution equation for $\Gamma_k[\psi, A, \bar A]$ obtains
a contribution from the variation of the infrared cutoff of the ghost
fields as given by
\bear
\frac{\partial}{\partial t}\Gamma_k[\psi,A,\bar A]&=&\frac{1}{2}\Tr \left\{
\left(\frac{\partial}{\partial t} R_k^{(A)}\right)\left(\Gamma_k^{(2)}
+R_k\right)^{-1}\right\}\nonumber\\
&&-\Tr \left\{\left(\frac{\partial}{\partial t}  
R_k^{(\psi)}\right)\left(\Gamma_k^{(2)}+R_k\right)^{-1}\right\}-
\varepsilon_k
\label{3.16}\\
\varepsilon_k&=&\Tr\left\{\left(\frac{\partial}{\partial  
t}R_k^{(gh)}\right)\left(\Gamma_k^{(gh)(2)}+R_k^{(gh)}\right)^{-1}\right\}
\label{3.17}
\ear
Here $\Gamma^{(2)}_k+R_k$ in (\ref{3.16}) is the matrix of second functional
derivatives of $\Gamma_k+\Delta_k^{(\psi)} S+\Delta_k^{(A)}S$ with respect
to $\psi$ and $A$ at fixed $\bar A$.
As compared to the more symmetric form of the flow
equation (A.12) we have combined here
similar pieces in the quark and ghost sector.
One should remember, however, that the
fermionic part $(\Gamma_k^{(2)}+R_k)^{-1}$ is a
submatrix of a larger matrix containing also
$\bar\psi\psi$ entries. For the derivation
of eq. (\ref{3.16})  we have exploited that the
matrix of second functional derivatives of  $(\Gamma_k+\Delta_kS)
[\psi,\xi, A, \bar A]$ is block diagonal in the $(\psi, A)$ and $\xi$
components for $\xi=0$.  The ghost dependence of $\Gamma_k[\psi,\xi,
A,\bar A]$
appears in the evolution equation for $\Gamma_k [\psi, A,\bar A]$ only
through
the term $\varepsilon_k$ which involves the second functional derivative
with
respect to the ghost fields $\Gamma_k^{(gh)(2)}$, which is evaluated at
$\bar\xi=\xi=0$ and may depend on $\psi,A,\bar A$.
As a consequence of local gauge invariance the average action $\Gamma_k$
must obey anomalous Slavnov-Taylor identities which are displayed
in the appendix. They constrain, in particular, the ghost dependence
of $\Gamma_k$.
We will not  pay much  attention to the detailed form of
$\Gamma_k^{(gh)(2)}$ in the present paper and approximate  it by its
``classical'' value (cf. (\ref{3.6}))
\be\label{3.18}
\Gamma_k^{(gh)(2)}=-D^\mu[\bar A] D_\mu[A].
\ee
or a slight generalization thereof (cf. eq. (4.8)).
In order to complete the formal setup  of our investigation we need to
specify
the operator ${\cal D}$ in eq. (\ref{3.10}).  A good choice is
\be\label{3.19}
{\cal D}[\bar A]=\Gamma_k^{(A)(2)}[\bar A]
\ee
where $\Gamma_k^{(A)(2)}$ is the second functional derivative of
$\Gamma_k[\psi,A,\bar A]$ with respect to $A$ for fixed $\bar A$ and
$\psi=0$, evaluated at the point $A=\bar A$. As in previous formulations
\cite{Reu}, the effective average action $\Gamma_k[\psi,A,\bar A]$ is gauge
invariant with respect to  simultaneous  gauge transformations of $\psi,
A$ and $\bar A$.

We can now  apply the formalism of the last section in order to
``integrate out'' the gluon fields $A$.
The classical field equation, whose solution is $A_k$,
reads
\be\label{3.20}
\frac{\delta\tilde\Gamma_k[\psi,A,\bar A]}{\delta A^z_\mu(x)}_{|A=A_k}=0
\ee
where the derivative should be taken at fixed $\bar A$.
At this point $A_k$ becomes a functional of $\psi$ and $\bar A$.
For the purpose of the present paper we only consider the special
choice $\bar A=0$ and omit the argument $\bar A$ in the
following.
In this version $R_k^{(A)}$ becomes  a simple function of momenta.
Summarizing our adaptation of eq. (\ref{2.17}) for quarks and gluons
one obtains
\bear\label{3.23}
\frac{\partial}{\partial t}\Gamma_k[\psi]&=&-
\left(\Gamma^{(2)}_k[\psi]+R_k^{(\psi)}\right)^{-1\beta}
_{\ \ \  \beta'}\left\{\left(\frac{\partial R_k^{(\psi)}}
{\partial t}\right)^{\beta'}_{\ \beta}
-\frac{\partial A^*_{k\alpha'}}{\partial\bar\psi_{\beta'}}
\left(\frac{\partial R_k^{(A)}}{\partial t}\right)^{\alpha'}_{\ \alpha}
\frac{\partial A^\alpha_k}{\partial  \psi^\beta}\right\}\nonumber\\
&&+\frac{1}{2}\left(\Gamma^{(2)}_k[\psi,A=A_k[\psi]]+R_k^{(A)}
\right)^{-1\alpha}_{\phantom{-1}\  \alpha'}\left(\frac{\partial R_k^{(A)}}
{\partial t}\right)^{\alpha'}_{\ \alpha}\nonumber\\
&&+\frac{1}{2}A^*_{k\alpha'}\left(\frac{\partial R_k^{(A)}}
{\partial t}\right)^{\alpha'}_{\ \alpha} A^\alpha_k -
\varepsilon_k[\psi, A_k[\psi]]
\ear
The flow equation (\ref{3.23})
is the central equation of this paper.
In order to exploit this equation we need $A_k[\psi],\
\Gamma^{(2)}_k[\psi,A_k]$ and $\varepsilon_k[\psi,A_k]$.

Our aim is the solution of the flow equation for $k\to0$,
starting at some high scale $k_0$ where $\Gamma_{k_0}[\psi]$
can be reliably computed by solving the field equation
for $A_{k_0}$ in a perturbative context. As $k$ decreases, we
gradually explore the non-perturbative regime and the
full quantum-effective action obtains for $k=0$. Obviously, such
a program is only feasible with approximations that truncate
the most general form of $\Gamma_k$. The truncation used in
the next two sections concerns the three-gluon and four-gluon
vertices which are approximated by a momentum-independent, but
$k$-dependent running gauge coupling.  A similar truncation
is used for the ghost contribution. In sect. 5 we truncate, in
addition, the most general form of the four- and six-fermion
interactions.

We are interested in the evolution
of the two- and four-point functions for the quarks. The respective
flow equations for these quantities obtain by taking the second and
fourth functional derivative of eq. (\ref{3.23}) at $\psi=\bar\psi=0$.
We label the
different contributions on the r.h.s. of eq. (\ref{3.23}) by
\be\label{4.18}
\frac{\partial}{\partial t}\Gamma_k[\psi]=-\gamma_\psi+\gamma_{A\psi}
+\gamma_A+\gamma_c-\epsilon\ee
and discuss them separately.
The first term
\be\label{4.19}
\gamma_\psi=Tr\left\{\left(\Gamma_k^{(2)}[\psi]+R^{(\psi)}_k\right)^{-1}
\frac{\partial}{\partial t}R_k^{(\psi)}\right\}\ee
is the standard contribution of a pure fermionic theory.
The remaining terms $\gamma_{A\psi},\gamma_A$ and $\gamma_c$
involve $R_k^{(A)}$ and reflect the contributions from gluons, whereas
$\epsilon$ gives the ghost contribution.
The term
\be\label{4.20}
\gamma_A=\frac{1}{2}Tr\left\{\left(\Gamma_k^{(2)}+R_k^{(A)}\right)^{-1}
\frac{\partial}{\partial t}R_k^{(A)}\right\}\ee
involves only a trace over gluonic degrees of freedom and
accounts for the contribution of gluon fluctuations around the
$\psi$-dependent
classical solution. The contribution of $\gamma_A$ to the
fermionic two- and four-point functions is given by the dependence
of $\Gamma_k^{(2)}[\psi,A_k[\psi],0]_{\ \alpha'}^\alpha$ on $\psi$.
Relevant contributions to $\gamma_A$ therefore arise from terms in
$\Gamma_k[\psi,A,0]$ which are either quadratic in $A$ and also depend
on $\psi$ or are cubic or higher-order in $A$.

We use a truncation where the first sort
of terms is absent and no relevant contribution to $\gamma_A$
would be present for an abelian gauge theory. For nonabelian
gauge theories we get contributions
>from the three- and four-gluon vertices in $\Gamma_k[\psi,A]$.
We approximate here these vertices by the (standard) lowest order
expressions which are obtained from functional derivatives of $F_{\mu\nu}
F^{\mu\nu}$. More precisely, we use on the r.h.s. of the flow
equation
\be\label{4.21}
\frac{\delta}{\delta A^\rho_w(y)}
\left(\Gamma_k^{(2)}[\psi,A,\bar A=0]\right)^{y\mu}_{\nu z}
(x,x')=\frac{\delta}{\delta A^\rho_w(y)}
\left(\tilde{\cal
D}[A]\right)^{y\mu}_{\nu z}\delta(x-x')\ee
with
\be\label{4.22}
\tilde Z_F^{-1}\left(\tilde{\cal D}[A]\right)^{y\mu}_{\nu z}
=-(\tilde D^2[A])^y_{\ z}\delta^\mu_\nu+2i\tilde
g\left(T_w\right)^y_{\ z}F^{w\mu}_{\ \nu}+(\tilde D_\nu[A]
\tilde D^\mu[A])^y_{\ z}\ee
Here $\tilde D_\mu[A]=\partial_\mu-i\tilde g A^z_\mu T_z$
represents the covariant derivative in
the adjoint representation with gauge coupling $\tilde g$
and $F_{\mu\nu}$ is the nonabelian field strength
associated to the gauge field $A_\mu$. In this truncation
the three- and four-gluon vertices are parametrized by two
running parameters $\tilde g(k)$  and $\tilde Z_F(k)$.
The running renormalized gauge coupling $g_k$ is related
to them by
\be\label{2.39}
g_k^2=\tilde g^2\tilde Z_F^{-1}\ee
We observe that $\Gamma_k[\psi,A,0]$ is
invariant under global gauge transformations of $\psi$ and $A$.
The expression
for $\gamma_A$ does not explicitly depend on $\psi$ in our
truncation and $\gamma_A[\psi=0,A_k]$ or $\Gamma_k[0,A_k,0]$
cannot contain a term linear
in $A_k$. There is therefore no contribution from $\gamma_A$ to the
flow equation of the fermionic two-point function.
An estimate of the contribution to the
four-quark interaction from
$\gamma_A$ therefore amounts
to a computation of the gluon contribution to the evolution of the term
quadratic in $A$ in $\Gamma_k[\psi=0,A,\bar A=0]$.

Similarly the ghost contribution $\epsilon$ is (with the
approximation (\ref{3.18})) only a functional of $A$, containing terms
quadratic in A (and higher orders).
We also observe that
$\gamma_A$ and $\epsilon$ only
account for the gluon and ghost contributions to  the effective
gluon propagator, whereas  the contribution from quark loops is implicitly
contained in $\gamma_\psi$. We note that the latter is not
distinguished any more from any other fermionic contributions, as, for
example, from an explicit four-quark interaction in ${\cal L}_k[\psi]$.
The contribution
\be\label{2.40}
\gamma_{c}=\frac{1}{2}A_{k\alpha'}^*\left(
\frac{\partial R_k^{(A)}}{\partial t}
\right)^{\alpha'}_{\ \alpha}\ A^\alpha_k\ee
describes the effect of the ``classical'' change in the
infrared cutoff as $k$ is lowered. It is quadratic
in the classical solution $A_k[\psi]$ and therefore gives
a contribution to the fermionic four-point function. Finally
the piece $\gamma_{A\psi}$ involves the explicit $\psi$-dependence
of the classical solution $\sim \partial A_k/\partial\psi$.
It contributes to the running of the fermionic two- and four-point
functions.

\section{Heavy quark approximation}
\setcounter{equation}{0}

In the limit of infinitely large quark masses our formalism simplifies
considerably. For euclidean momenta we
can omit in eq. (\ref{3.23}) the terms involving the
inverse fermion propagator $(\Gamma_k^{(2)}[\psi]+R_k^{(\psi)})^{-1}$
since their contribution is suppressed by inverse powers of the
quark masses.
In the language of the last section this results in
$\gamma_\psi=0,\gamma_{A\psi}
=0$. The remaining computation amounts to an investigation of the
pure gluon
theory with static quarks. This is done most easily in the language
where the gluon fields are kept explicitly and the relevant effective
action is $\Gamma_k[\psi,A,\bar A=0]$.
If one wants to extract the effective four-quark interaction,
one needs the $k$-dependent
effective gluon
propagator and the effective vertex $\bar\psi\psi A$.
We first describe for arbitrary quark masses the general
framework how an effective four-quark interaction
obtains from ``gluon
exchange'' in the formulation where both quark and gluon
degrees of freedom
are kept explicitly. We then specialize to the heavy quark limit.

Let us consider in $\Gamma_k[\psi,A,\bar A]$ the term quadratic
in $A$
\be\label{5.1}
\Gamma_{k,2}^{(A)}=\frac{1}{2}\int\frac{d^4q}{(2\pi)^4}A^\nu_y(-q)
(\Gamma_k^{(2)}[\psi=0,A=0,\bar A=0])^{y\mu}_{\nu z}A^z_\mu(q)\ee
and parametrize
the most general inverse gluon propagator by
\be\label{5.2}
\left(\Gamma_k^{(2)}[\psi=0,A=0,\bar A=0]\right)^{y\mu}_{\nu z}
(q)=(G_A(q)\delta^\mu_\nu+H_A(q)q_\nu q^\mu)\delta^y_z\ee
For the quark-gluon vertex
\be\label{5.3}
\Gamma_k^{(\bar\psi\psi A)}=\int\frac{d^4p}{(2\pi)^4}\frac{d^4q}
{(2\pi)^4}\bar\psi^i_a(p)G_\psi(p,q)
\gamma^\mu(T_z)^{\ j}_i\psi^a_j(p+q)A^z_\mu(-q)\ee
we make the approximation that $G_\psi$ is a simple function not
involving Dirac matrices.
Knowledge of $G_A, H_A$ and $G_\psi$ permits to compute the classical
solution  $A_k$ in order $\bar\psi\psi$
\be\label{5.4}
(A_k^{(0)}(q))^\nu_z=-S^\nu_\mu(q)\int\frac{d^4p}{(2\pi)^4}
G_\psi(p,q)
\bar\psi_a^i(p)\gamma^\mu(T_z)_i^{\ j}\psi^a_j(p+q)\ee
Here $S=(\Gamma_k^{(2)}[0]+R_k)^{-1}$ is the gluon propagator in presence
of the infrared cutoff
\be\label{5.5}
(R_k^{(A)})^{y\mu}_{\nu z}(q)=(R_k(q)\delta^\mu_\nu+
\tilde R_k(q)q_\nu q^\mu)\delta^y_z\ee
and reads
\bear\label{5.6}
&&S^\nu_\mu(q)=(G_A(q)+R_k(q))^{-1}\left\lbrace\delta^
\nu_\mu-q^\nu q_\mu(H_A(q)+\tilde R_k(q))\cdot\right.\nonumber\\
&&\left.[G_A(q)+R_k(q)+q^2(H_A(q)+\tilde R_k(q))]^{-1}
\right\rbrace\ear
Inserting the classical solution into (\ref{5.2}) and (\ref{5.3}) and
accounting for the term $\Delta_k^{(A)}S$ (2.22) we find
the effective quark four point function
\bear\label{5.7}
&&\Gamma_{k,4}^{(\psi)}=-\frac{1}{2}\int\frac{d^4p}{(2\pi)^4}
\frac{d^4p'}{(2\pi)^4}\frac{d^4q}{(2\pi)^4}S^\nu_\mu(q)
G_\psi(p,q)G_\psi(p',-q)
\nonumber\\
&&\lbrace\bar\psi^i_a(p)\gamma^\mu(T_z)_i^{\ j}\psi^a_j(p+q)
\rbrace\lbrace
\bar\psi^k_b(p')\gamma_\nu(T^z)_k^{\ l}\psi^b_l(p'-q)\rbrace
\nonumber\\
&&=-\frac{1}{2}\int\frac{d^4p_1}{(2\pi)^4}...\frac{d^4p_4}
{(2\pi)^4}(2\pi)^4
\delta(p_1+p_2-p_3-p_4)\cdot\nonumber\\
&&\left\lbrace F_1(p_1,p_2,p_3,p_4){\cal M}(p_1,p_2,p_3,p_4)
+F_2(p_1,p_2,p_3,p_4){\cal N}(p_1,p_2,p_3,p_4)\right\rbrace\ear
with
\bear\label{5.8}
F_1&=&G_\psi(-p_1,p_1-p_3)G_\psi(p_4,p_2-p_4)(G_A(p_1-p_3)+
R_k(p_1-p_3))^{-1}
\nonumber\\
F_2&=&G_\psi(-p_1,p_1-p_3)G_\psi(p_4,p_2-p_4)(H_A(p_1-p_3)+
\tilde R_k(p_1-p_3))\nonumber\\
&&(G_A(p_1-p_3)+R_k(p_1-p_3))^{-1}[G_A(p_1-p_3)+R_k(p_1-p_3)\nonumber\\
&&+(p_1-p_3)^2(H_A(p_1-p_3)+\tilde R_k(p_1-p_3))]^{-1}\ear
and
\bear\label{5.9}
{\cal N}(p_1,p_2,p_3,p_4)&=&\lbrace\bar\psi^i_a(-p_1)(p\llap/_1-p\llap/_3)
(T_z)_i^{\ j}\psi^a_j(-p_3)\rbrace\nonumber\\
&&\lbrace\bar\psi^k_b(p_4)(p\llap/_2-p\llap/_4)
(T^z)_k^{\ l}\psi^b_l(p_2)\rbrace\ear
\be\label{4.11}
{\cal M}(p_1,p_2,p_3,p_4)=\left\{\bar\psi^i_a(-p_1)\gamma^\mu(T^z)_i^{\  
j}\psi_j^a(-p_3)\right\}\left\{\bar\psi^k_b(p_4)\gamma_\mu(T_z)_k
^{\ \ell}\psi_\ell^b(p_2)\right\}\ee
The curled brackets indicate  contractions over  not explicitly written
indices (here spinor indices), $i,j,k,\ell=1...N_c$ are the colour
indices and $a,b=1...N_f$ the flavour indices of the quarks.
By an appropriate Fierz transformation and using the identity
\be\label{4.12}
(T^z)_i^{\ j}(T_z)_k^{\ \ell}=\frac{1}{2}\delta^\ell_i\delta^j_k-
\frac{1}{2N_c}\delta^j_i\delta^\ell_k
\ee
we can split ${\cal M}$ into three terms \cite{Ell}
\bear
{\cal M}&=&{\cal M}_\sigma+{\cal M}_\rho+{\cal M}_p\label{4.13}\\
{\cal M}_\sigma&=&-\frac{1}{2}\left\{\bar\psi^i_a(-p_1)\psi_i^b(p_2)\right\}
\left\{\bar\psi^j_b(p_4)\psi_j^a(-p_3)\right\}\nonumber\\
&&+\frac{1}{2}\left\{\bar\psi^i_a(-p_1)\gamma^5\psi_i^b(p_2)\right\}\left\{
\bar\psi^j_b(p_4)\gamma^5\psi_j^a(-p_3)\right\}\label{4.14}\\
{\cal M}_\rho&=&\frac{1}{4}\left\{\bar\psi^i_a(-p_1)\gamma_\mu\psi_i^b(p_2
\right\}\left\{\bar\psi^j_b(p_4)\gamma^\mu\psi_j^a(-p_3)\right\}\nonumber\\
&&+\frac{1}{4}\left\{\bar\psi^i_a(-p_1)\gamma_\mu\gamma^5
\psi_i^b(p_2)\right\}
\left\{\bar\psi^j_b(p_4)\gamma^\mu\gamma^5\psi_j^a(-p_3)\right\}
\label{4.15}\\
{\cal M}_p&=&-\frac{1}{2N_c}\left\{\bar\psi^i_a(-p_1)
\gamma_\mu\psi_i^a(-p_3)
\right\}\left\{\bar\psi^j_b(p_4)\gamma^\mu\psi_j^b(p_2)\right\}
\label{4.16}\ear
In terms of the Lorentz invariants
\bear\label{4.17}
s&=&(p_1+p_2)^2=(p_3+p_4)^2\nonumber\\
t&=&(p_1-p_3)^2=(p_2-p_4)^2\ear
we recognize that the quantum numbers of the fermion bilinears in ${\cal M}_
\sigma$ correspond to colour singlet, flavour non-singlet scalars in the
$s$-channel and similarly for spin-one mesons for ${\cal M}_\rho$. In
analogy to ref. \cite{Ell} we associate these terms with the scalar mesons of
the linear $\sigma$-model and with the $\rho$-mesons. The bilinears in the
last term ${\cal M}_p$ correspond to a colour and flavour singlet spin-one
boson in the $t$-channel. These are the quantum numbers of the pomeron.
We observe that in the heavy quark approximation
where $\Gamma^{(\psi)}_{k,4}$ arises only from ``gluon
exchange'', the coefficients of the quark
interactions in the $\sigma,\rho$ and pomeron channel (\ref{4.13}) are all
given by the same function $F_1$.

The general
quark bilinear is conveniently parametrized by the real functions
$Z_\psi(q)$ and $\bar m_a(q)$
\be\label{5.10}
\Gamma_{k,2}^{(\psi)}=\sum_a\int\frac{d^4q}{(2\pi)^4}\bar\psi_a^i(q)
(Z_\psi
(q)\gamma^\mu q_\mu+\bar m_a(q)\gamma^5)\psi^a_i(q)\ee
The $k$-dependence of the functions $G_A,H_A,G_\psi,Z_\psi$ and $\bar m_a$
relevant for the two- and four-point functions for the quarks can now
be studied using the evolution equation (\ref{3.16}) for $\Gamma_k
[\psi,A,\bar
A=0]$. In the truncation where only the terms (\ref{5.2}),(\ref{5.3})
and (\ref{5.10}) are kept, it is easy to see that the contributons to the
$k$-dependence of $G_\psi,Z_\psi$ and $\bar m_a$ all involve quark
propagators. In the heavy quark limit they can therefore be neglected
for euclidean external momenta. Only
the $k$-dependence of $\Gamma_{k,2}^{(A)}$ needs to be considered.
For $Z_\psi$ and $G_\psi$ we may take appropriate momentum-independent
``short-distance couplings''
\bear\label{5.11}
Z_\psi(q)&=&1\nonumber\\
G_\psi(p,q)&=&\tilde Z_F^{\frac{1}{2}}(m_\psi)g(m_\psi)
=\tilde g(m_\psi)\ear
with renormalized gauge coupling $g$ taken at the scale $k=m_\psi$
and $m_\psi$ the heavy  quark mass. We also may identify $k=m_\psi$
with the
``ultraviolet cutoff'' or the scale where the initial values for the flow
equation are specified, i.e.
\bear\label{5.12}
\tilde Z_F(m_\psi)&=&1\nonumber\\
G_A(q;k=m_\psi)&=&q^2\ear
Solving the flow equation for $G_A(q)$ for $k\to0$
yields the effective four-quark interactions for momenta
much smaller than the quark mass.
For $\alpha_R=0$ (see next section) the effective four-quark interaction
is fully determined by
\bear\label{5.13}
&&\Gamma_{0,4}^{(\psi)}=-\frac{1}{2}g^2(m_\psi)\int\frac{d^4p_1}
{(2\pi)^4}...
\frac{d^4p_4}{(2\pi)^4}(2\pi)^4\delta(p_1+p_2-p_3-p_4)\nonumber\\
&&\lim_{k\to0}G^{-1}_A(p_1-p_3)\left\lbrace{\cal M}(p_1,p_2,p_3,p_4)
+\frac{1}{(p_1-p_3)^2}{\cal N}(p_1,p_2,p_3,p_4)\right\rbrace\ear

We finally should mention that the heavy quark
potential or the scattering amplitude for heavy quarks cannot be
extracted directly from the four-point function at small momenta
$p^2\ll m_\psi^2$. For these purposes the momenta appearing in
${\cal M}$ should be taken on-shell, i.e. $p_1^2=p_2^2=p_3^2
=p^2_4=-m_\psi^2$. Their size is therefore not small as compared
to $m_\psi^2$. For on-shell momenta the vertex function
$G_\psi(p,q)$ becomes a function of $q^2$. The flow of
$G_\psi(q)$ does not vanish for $k^2\ll m^2_\psi$ and we need
to supplement the computation of the $k$-dependence of
$G_A(q)$ by a corresponding one of $G_\psi(q)$. The heavy quark
pontential can then be extracted as the (three-dimensional)
Fourier transform of $F(q)=G_\psi^2(q)G_A^{-1}(q)$.

\section{Scale dependence of the gluon propagator}
\setcounter{equation}{0}

In this section we compute a flow equation for
the scale dependence of the effective
gluon propagator in the pure Yang-Mills theory. As explained in the last
section, this is the central piece which determines the heavy quark
interactions. We will derive the flow equation for $G_A$ in the
approximation that the vertices can be extracted from a term $\frac{1}{4}
\tilde Z_FF^z_{\mu\nu}F^{\mu\nu}_z$ in the average action, with running
wave function renormalization
$\tilde Z_F$ and running gauge coupling. The resulting functional
form (\ref{6.9}) of the r.h.s. of the flow equation corresponds
to $\gamma_A
-\epsilon$ and is valid beyond the heavy quark approximation.
The second part of this section discusses qualitative properties of the
solution of the flow equations. This part is valid only for the pure
Yang-Mills theory or for the heavy quark theory.
For the derivation of the
evolution equation we keep the most general gluon
propagator. The only truncation
concerns the momentum dependence of the effective three-gluon
and four-gluon
vertices and the ghost sector.
One finds that the $k$-dependence of the functions
$G_A(q)$ and $H_A(q)$ is governed by the flow equations
\bear\label{6.1}
&&\frac{\partial}{\partial t}G_A(q)=N_cg^2_k\tilde Z_F\int\frac{d^4q'}
{(2\pi)^4}\tilde\partial_t\Biggl\lbrace
\left(G_A(q')+R_k(q')\right)^{-1}\tilde Z_F\left[3-\frac{3}{4}
b(q')\right]\nonumber\\
&&-\frac{1}{2}\left(G_A(q')+R_k(q')\right)^{-1}
\left(G_A(q+q')+R_k(q+q')\right)^{-1}\tilde Z^2_F\nonumber\\
&&\left[5q^2+2(qq')+\frac{16}{3}{q'}^2-\frac{10}{3}\frac{(qq')^2}{q^2}
\right.\nonumber\\
&&-\frac{2}{3}b(q')\left(2{q'}^2+10(qq')+q^2+11\frac{(qq')^2}{{q'}^2}+2
\frac{(qq')^3}{q^2{q'}^2}+\frac{(qq')^2}{q^2}\right)\nonumber\\
&&\left.+\frac{1}{3}b(q')b(q+q')\frac{q^2}{(q+q')^2}\left(q^2-
\frac{(qq')^2}{{q'}^2}\right)\right]
\nonumber\\
&&+\frac{1}{3}P^{-1}_{\rm gh}(q')P^{-1}_{\rm gh}(q+q')
\left[{q'}^2-\frac{(qq')^2}{q^2}\right]\Biggr\rbrace\ear
and
\bear\label{6.2}
&&\frac{\partial}{\partial t}H_A(q)=-\frac{1}{2}N_cg^2_k\tilde  
Z_F\int\frac{d^4q'}{(2\pi)^4}\tilde\partial_t\nonumber\\
&&\Biggl\lbrace\left(G_A(q')+R_k(q')\right)^{-1} \left(G_A(q+q')+
R_k(q+q')\right)^{-1}\tilde Z_F^2\nonumber\\
&&\left[\frac{40}{3}\frac{(qq')^2}{(q^2)^2}-
\frac{10}{3}\frac{{q'}^2}{q^2}+10\frac{(qq')}{q^2}-2
\right.\nonumber\\
&&-2b(q')\left(\frac{2}{3}-\frac{4}{3}\frac{(qq')}{q^2}-\frac{14}{3}
\frac{(qq')^2}{q^2{q'}^2}+\frac{1}{3}\frac{{q'}^2}{q^2}-\frac{4}{3}
\frac{(qq')^2}{(q^2)^2}-\frac{8}{3}\frac{(qq')^3}{(q^2)^2{q'}^2}
\right)\nonumber\\
&&\left.+\frac{1}{3}b(q')b(q+q')\frac{1}{(q+q')^2}
\left(\frac{(qq')^2}{{q'}^2}
-q^2\right)\right]\nonumber\\
&&-\frac{2}{3}P^{-1}_{\rm gh}(q')P^{-1}_{\rm gh}(q+q')
\left[3\frac{(qq')}{q^2}+4\frac{(qq')^2}{(q^2)^2}-
\frac{{q'}^2}{q^2}\right]\Biggr\rbrace\ear
Here we use
\be\label{6.3}
b(q)=\frac{(H_A(q)+\tilde R_k(q))q^2}{G_A(q)+R_k(q)+(H_A(q)
+\tilde R_k(q))q^2}\ee
and note that the partial derivative $\tilde\partial_t$
acts only on the explicit infrared cutoff terms $R_k$ and $\tilde R_k$.
The parts involving the effective ghost propagator $P^{-1}_{\rm gh}$
arise from the ghost contribution $\sim-\epsilon$.
After performing the $q'$ integration the evolution equations (\ref{6.1}),
(\ref{6.2}) can be interpreted as two coupled nonlinear partial
differential
equations for the functions $G_A,H_A$ which depend on two variables
$k$ and $q^2$. They describe the scale dependence of the gluon
propagator in the approximation where both the three-gluon and
the four-gluon vertex are given by a single renormalized gauge coupling
$g_k$ and similarly for the ghost gluon vertex.

Even in this approximation the flow equations are lengthy and difficult
to solve. A simplification occurs if we take for the gauge-fixing term
\be\label{6.4}
S_{\rm gf}=\frac{1}{2\alpha}\int d^4x(\partial^\mu A^z_\mu)^2\ee
the gauge parameter $\alpha\to0$. In this limit $H_A$ diverges
$\sim\frac{1}
{\alpha}$ and $b(q)$ approaches one. We may define a $k$-dependent
renormalized gauge fixing parameter $\alpha_R$ by\footnote{We choose
$\tilde R_kq^2=\left(\frac{1}{\alpha_R}-1\right)R_k$ such that
$b(q)=1+0(\alpha_R)$ for all values of $q^2$.}
\be\label{6.5}
H_A(0)=\left(\frac{1}{\alpha_R}-1\right)\tilde Z_F\ee
with $\lim_{k\to\infty}\tilde Z_F=1$ and $\lim_{k\to\infty}
\alpha_R=\alpha$. The evolution equation for $\alpha_R$ follows from
(\ref{6.2})
\be\label{6.6}
\frac{\partial}{\partial t}\alpha_R=\frac{\partial \ln
\tilde Z_F}{\partial t}
(\alpha_R-\alpha_R^2)-\alpha_R^2\tilde Z^{-1}_F\frac{\partial}{\partial t}
H_A(0)\ee
where we note that $\frac{\partial}{\partial t}
H_A(q)$ has a well defined
limit for $q^2\to0$. We conclude that $\alpha_R=0$ is a fixpoint which is
infrared stable for $\partial \ln\tilde Z_F/\partial t>0$.
The approximation $\alpha_R\to0$ remains therefore stable in the course
of the evolution.
We further observe that for $\alpha_R\to0$ the function $F_2$ defined
in eq. (\ref{5.8}) equals $F_1/(p_1-p_3)^2$ and only the function
$G_A(q)$ determines the effective four-quark interaction. The flow
equation for $G_A$ can be written in a more compact form using
\be\label{6.7}
P_A(q)=\tilde Z^{-1}_F(G_A(q)+R_k(q))\ee
We also approximate the ghost part of $\tilde \Gamma_k^{(2)}$ by
\be\label{6.8}
P_{\rm gh}(q)=P_A(q),\qquad \tilde\partial_tP_{\rm gh}(q)
= \tilde\partial_tP_A(q)\ee
This yields for $\alpha_R\to0$
\bear\label{6.9}
&&\frac{\partial}{\partial t}G_A(q)=N_cg^2_k\tilde Z_F\int\frac{d^4q'}
{(2\pi)^4} \tilde\partial_t\Biggl\lbrace
\frac{9}{4}P_A^{-1}(q')\nonumber\\
&&-\frac{1}{6}P^{-1}_A(q')P^{-1}_A(q+q')\Biggl[13q^2-14(qq')+10{q'}^2
\nonumber\\
&&-10\frac{(qq')^2}{q^2}-22\frac{(qq')^2}{{q'}^2}-4\frac{(qq')^3}
{q^2{q'}^2}+\frac{q^2}{{q'}^2}\frac{q^2{q'}^2-(qq')^2}
{(q+q')^2}\Biggr]\Biggr\rbrace\ear

The evolution equation (\ref{6.9}) is the central equation of this section.
It is a partial nonlinear differential equation for $G_A(q^2;k)$ which
can be solved numerically. We observe that the
r.h.s. of the
flow equation (\ref{6.9}) involves not only $G_A(q)$ but also
the renormalized gauge coupling
$g_k$ and the gluon wave function renormalization constant $\tilde Z_F$.
The precise definition of these $k$-dependent constants
is a somewhat subtle issue. We could define $\tilde g$
and $\tilde Z_F$ (and in consequence $g_k$) in terms of the effective
three-gluon vertex $\sim \tilde g\tilde Z_FA^2\partial A$ and four-gluon
vertex $\sim \tilde g^2\tilde Z_FA^4$ which enter the approximation for
$\Gamma^{(2)}_k$ used in (2.38). In this way $\tilde g$ and $\tilde
Z_F$ are expressed in terms of third and fourth functional
derivatives of $\Gamma_k[\psi=0,A,\bar A=0]$ evaluated at $A=0$ and
projected on the
appropriate index structures. One also has to choose appropriate momenta
for the external legs for the effective vertices, as for example the limit
where all momenta approach zero. Evolution equations for the scale
dependence of $\tilde Z_F, \tilde g$ and $g_k$
could then be computed from appropriate
functional derivatives of the flow equation (\ref{3.16}). We will
use here a
simplification and approximate $\tilde Z_F$ by the coefficient
of the $q^2$
term in $G_A$. More precisely, we expand for small $q^2$
\be\label{6.10}
G_A(q)=\bar m^2_A+G^{(1)}_Aq^2+G_A^{(2)}(q^2)^2+...\ee
and identify $\tilde Z_F$ with $G_A^{(1)}$. There is an
obvious limitation
to this approximation since $G_A^{(1)}$ may turn negative for small $k$
and any reasonable choice of a wave function renormalization
requires positive $\tilde Z_F$. For the
flow equation for the renormalized gauge coupling $g_k$ we rely
on the fact that the first two coefficients of the $\beta$ function
\be\label{6.11}
\frac{\partial g^2_k}{\partial t}=\beta_{g^2}=-c_1\frac{g^4_k}
{16\pi^2}-c_2\frac{g^6_k}{(16\pi^2)^2}-...\ee
are universal
\bear\label{6.12}
c_1=\frac{22N_c}{3}\qquad c_2=\frac{204}{9}N^2_c\ear
In the region of large $g_k$ we may also use nonperturbative estimates
of $\beta_{g^2}$ derived by different methods \cite{Reu}. An ansatz for
$\beta_{g^2}$ combined with an estimate of $\tilde\eta_F$ fixes also
the evolution of $\tilde g^2$ and provides all information needed for
a numerical solution of the flow equation.

We concentrate first on an analytic discussion of a few
prominent features of the solution of equation (\ref{6.9}). The
evolution
equations for the mass term $\bar m^2_A$ and for $G_A^{(1)}$ are
easily derived by expanding the r.h.s.
of eq. (\ref{6.9}) in powers of $q$.
One obtains
\be\label{6.13}
\frac{\partial}{\partial t}\bar m^2_A=N_cg^2_k\tilde Z_F\int\frac{d^4q'}
{(2\pi)^4} \tilde\partial_t\left\lbrace\frac{9}{4}
P_A^{-1}(q')-\frac{5}{4}{q'}^2P^{-2}_A(q')\right\rbrace\ee
and the renormalized dimensionless mass term
\begin{equation}\label{6.14}
\tilde m^2_A=\bar m^2_A\tilde Z_F^{-1}k^{-2}\end{equation}
therefore obeys
\begin{equation}\label{6.15}
\frac{\partial}{\partial t}\tilde m^2_A=(-2+\tilde\eta_F)\tilde m^2_A
-\frac{N_c}{8\pi^2}g^2_k\left(\frac{9}{4}l_{A,1}^4-\frac{5}{4}l^6_{A,2}
\right)
\end{equation}
Here we have defined the integrals, with $x={q'}^2$,
\be\label{6.16}
l^d_{A,n}=-\frac{1}{2}k^{2n-d}\int^\infty_0dxx^{\frac{d}{2}-1}
 \tilde\partial_tP^{-n}_A(x)\ee
The evolution of the ration $\tilde m^2_A/g^2_k$
\be\label{6.17}
\frac{\partial}{\partial t}\left(\frac{\tilde m^2_A}{g^2_k}\right)
=-\frac{N_c}{8\pi^2}\left(\frac{9}{4}l^4_{A,1}
-\frac{5}{4}l^6_{A,2}\right)-\left(2+\frac{\beta_{g^2}}{g^2_k}-\tilde\eta
_F\right)\frac{\tilde m^2_A}{g_k^2}\ee
is characterized for small $g^2_k$ by an approximate infrared
unstable fixpoint
\be\label{6.18}
\frac{\tilde m^2_A}{g^2_k}=-\frac{3N_c}{128\pi^2}\ee
Indeed, we can use for small $g^2_k$ the lowest order expressions
\bear\label{6.19}
G_A(q)&=&\tilde Z_Fq^2\nonumber\\
\tilde\partial_tP_A(q)&=&\frac{\partial}{\partial t}
P(q)\ear
with
\be\label{6.20}
P(q)=q^2+\tilde Z_F^{-1}R_k(q)=\frac{q^2}{1-\exp\left(-\frac{q^2}{k^2}
\right)}\ee
such that $l^4_{A,1}=l^4_1=1, l^6_{A,2}=l^6_2=\frac{3}{2}$. We also
neglect $\beta^2_g/g^2-\tilde\eta_F$ as compared to two. For small
$g^2_k$ the general identities for the dependence
of $\Gamma_k[\psi,A,\bar A]$
on the background field $\bar A$ \cite{Reu}, or, similarly, the generalized
Slavnov-Taylor identities \cite{Mar},\cite{Ell2}
imply that the $k$-dependent mass term
is indeed described by this fixpoint (cf. Appendix).
We conclude that for small $g_k$
the mass term
induces only a small correction in the momentum-independent part of
$P_A$
\bear\label{6.21}
P_A&\approx& P+\tilde Z_F^{-1}\bar m^2_A+O(q^4)\nonumber\\
&=&P-\frac{3N_c}{128\pi^2}g^2_kk^2+O(q^4)\nonumber\\
&=&k^2\left(1-\frac{3N_c}{128\pi^2}g^2_k\right)+O(q^2)\ear
The behaviour of $\bar m^2_A$ near the confinement scale where $g^2_k$
becomes large is more complicated and best described by evaluating
directly the relevant identities \cite{Reu}. It is conceivable
that a negative r.h.s. of (\ref{6.13}) drives $\bar m^2_A$ to a positive
value for $k\to0$, but we find this scenario not very likely.

The flow equation for $G_A^{(1)}$ can be written in the form
\bear\label{6.22}
\frac{\partial}{\partial t}G_A^{(1)}&=&\frac{N_c}{96\pi^2}g^2_k\tilde Z_F
(31l^4_{A,2}-5m^6_{A,4})\nonumber\\
&=&\frac{13}{3}N_cc_A\frac{\tilde g^2}{16\pi^2}\ear
where we have defined the integrals \be\label{6.23}
m^d_{A,n}=-\frac{1}{2}k^{2n-d-2}\int^\infty_0dxx^{\frac{d}{2}}
\tilde\partial_t\left(\dot P_A^2(x)P^{-n}_A(x)\right)\ee
with $\dot P_A=\frac{d}{dx}P_A$. We note that the r.h.s. of eq.
(\ref{6.22}) is
positive for positive $c_A$
\bear\label{6.24}
c_A&=&\frac{1}{26}(31l^4_{A,2}-5m^6_{A,4})\nonumber\\
&=&-\frac{31}{52}\int^\infty_0dx x\tilde\partial_t
\left\lbrace P^{-2}_A\left(1-\frac{5}{31}x^2\dot P^2_AP^{-2}_A\right)
\right\rbrace\ear
Let us first consider small values of the gauge coupling where we can
approximate $\tilde\partial_tP_A=\frac{\partial}{\partial t}
P$. In this limit we obtain $c_A=1$\footnote{We have also computed
$\frac{\partial}{\partial t}H_A(0)=-\frac{N_c}{48\pi^2}\tilde g^2
\left(8l^4_{A,2}+5m^6_{A,4}\right)$ and find for small $g^2$ that
$\frac{\partial}{\partial t}\left(H_A(0)+G^{(1)}_A\right)$ vanishes
as required by the Slavnov-Taylor identity for perturbative
trans\-ver\-sality. }. For the definition
\be\label{6.25}
\tilde Z_F=G^{(1)}_A\ee
and therefore finds
\be\label{6.26}
\tilde\eta_F=-\frac{\partial}{\partial t}\ln\tilde Z_F=
-\frac{13}{3}N_cc_A\frac{g^2_k}{16\pi^2}\ee
For a solution of (\ref{6.26}) it is convenient to compare
$\tilde\eta_F$ with
\be\label{6.27} \eta_F=\frac{\partial}{\partial t}\ln g^2_k=
\frac{\beta_{g^2}}{g^2_k}=-\frac{22}{3}N_cb_A\frac{g^2_k}{16\pi^2}\ee
where $b_A$ may depend on $g_k$ or $k$ and reflects the deviation
>from the one-loop $\beta$-function for which $b_A=1$. As long as the
$k$-dependence of the ratio $b_A/c_A$ can be neglected, one obtains
\be\label{6.28}
\frac{\tilde Z_F(k)}{\tilde Z_F(k_0)}=\left(\frac{g^2(k_0)}{g^2(k)}\right)
^\gamma,\
\gamma=\frac{13c_A}{22b_A}.\ee
In the one-loop approximation $\frac{1}{g^2(k)}$ decreases logarithmically
and reaches zero at the confinement scale $\Lambda_{conf}$. We conclude
that this is exactly the scale where $\tilde Z_F$ would vanish.
In this language the divergence of the renormalized gauge coupling
is actually entirely due to the vanishing of $\tilde Z_F$: From (\ref{6.28})
one obtains
\bear\label{6.29}
\frac{\tilde g^2(k)}{\tilde g^2(k_0)}&=&\left(\frac{g^2(k)}
{g^2(k_0)}\right)
^{1-\gamma}\nonumber\\
\frac{\tilde Z_F(k)\tilde g^2(k)}{\tilde Z_F(k_0)\tilde g^2(k_0)}&=&
\left(\frac{g^2(k)}{g^2(k_0)}\right)
^{1-2\gamma}\nonumber\\
\frac{\tilde Z_F(k)\tilde g(k)}{\tilde Z_F(k_0)\tilde g(k_0)}&=&
\left(\frac{g(k)}{g(k_0)}\right)
^{1-3\gamma}\ear
and for $\gamma>\frac{1}{2}$ (cf. (\ref{6.28})) both the
unrenormalized three-point vertex $\sim\tilde Z_F\tilde g$ and
four-point vertex $\sim \tilde Z_F\tilde g^2$ vanish at the confinement
scale.

As mentioned before we should not use (\ref{6.25}) for $k$
in the vicinity
of the confinement scale. We therefore propose \cite{Reu2}
to keep $\tilde Z_F$
independent of $k$ for $k<k_{np}$ where $k_{np}$ is defined
by
\be\label{6.30}
g^2_{k_{np}}=\frac{4\pi^2}{N_c}.\ee
The definition
\be\label{6.31}
\tilde Z_F=\left\{\begin{array}{lll}
G^{(1)}_A(k)& {\rm for}& g^2_k<\frac{4\pi^2}{N_c}\\
\tilde Z_F(k_{np})& {\rm for}& g^2_k\geq\frac{4\pi^2}{N_c}\\
\end{array}\right.\ee
allows to separate the issue of vanishing $G^{(1)}_A$ from the
choice of the infrared cutoff (i.e.$\tilde Z_F$).

The vanishing of the term $G_A^{(1)}q^2$ would have important
consequences for the behaviour of the propagator $\sim G^{-1}_A$. We
should therefore investigate if this feature is likely to survive
beyond the approximation of small gauge coupling. Turning back to
eq. (\ref{6.22}) we observe that $G_A^{(1)}$ can only remain positive
for $k\to0$ if $\tilde g^2$ turns to zero or if $c_A$ vanishes or
becomes negative. In view of eq. (\ref{6.29}) the first alternative
seems not very likely. (This probably generalizes if we go beyond
the approximation leading to (\ref{6.1}) and consider general
momentum-dependent three- and four-point functions. Then in eq.
(\ref{6.22}) $\tilde g^2$ has to be replaced by an appropriate
momentum-weighted
average of these vertices.) In order to investigate the sign of $c_A$
we write the integral (\ref{6.24})
explicitly as
\be\label{6.32}
c_A=\frac{31}{26}\int^\infty_0dx xP^{-3}_AS\left\lbrace
1-\frac{10}{31}x^2\left(\frac{\dot P_A}{P_A}\right)^2+\frac{5}
{31}x^2\frac{\dot P_A\dot S}{P_AS}\right\rbrace\ee
with
\bear\label{6.33}
S(x)=\tilde\partial_tP_A(x)&=&\frac{\partial}{\partial t}
P(x)-\tilde\eta_F(P(x)-x)\nonumber\\
&=&\left(2\frac{P(x)}{k^2}-\tilde\eta_F\right)(P(x)-x)\ear
The integral is dominated by the region $x\approx k^2$ and can turn
negative only if the bracket is negative in this region. As an
illustration we take
$P_A=P+m^2_A+\kappa x^2,m^2_A=\tilde Z_F^{-1}\bar m^2_A$. This implies
\be\label{6.34}
\frac{x\dot P_A}{P_A}=\frac{\dot P x+2\kappa x^2}{P+m^2_A+\kappa x^2}\ee
to be compared with
\be\label{6.35}
\frac{x\dot P}{P}=1-\frac{x}{k^2}\exp\left(-\frac{x}{k^2}\right)
\left(1-\exp\left(-\frac{x}{k^2}\right)\right)^{-1}=1-\frac{P-x}{k^2}\ee
and
\be\label{6.36}
\frac{x\dot S}{S}=\frac{x\dot P}{P}\left(1-\frac{\tilde\eta_F}{2}
\frac{k^2}{P}\right)^{-1}+1-\frac{P}{k^2}\ee
We conclude that negative $m^2_A$ (cf. (\ref{6.18})) and positive
$\kappa$ tend to lower $c_A$. A vanishing $c_A$ for large $\kappa$
cannot be excluded without a more detailed investigation.
On the other side,
if the integral relation
\be\label{6.37}
\int^\Lambda_{\Lambda_{conf}}\frac{dk}{k}\tilde g^2(k)c_A(k)
=\frac{48\pi^2}{13N_c}\ee
could be
fulfilled for $\Lambda_{conf}>0$, the coefficient $G_A^{(1)}$
would vanish
at the confinement scale $\Lambda_{conf}$
and presumably becomes negative
for $k<\Lambda_{conf}$. We emphasize that
a negative value of $\bar m^2_A$
or of $G_A^{(1)}$ for small $k$ would independently indicate
that the groundstate does not correspond to the perturbative ground
state $A_\mu=0$\ \cite{Sav},\cite{RGR},\cite{Reu2}.

Even before reaching the confinement scale, the gluon propagator
has to be modified: Whenever the term $G_A^{(1)}q^2$ becomes comparable
to $G_A^{(2)}(q^2)^2$ for $q^2\approx k^2$, i.e. for
$G_A^{(1)}\approx G_A^{(2)}k^2$, the gluon propagator cannot be
approximated any more by the inverse of $q^2$! As an example
for a plausible form one may consider
\be\label{6.38}
G_A(q)\approx \bar m^2_A+G^{(1)}_Aq^2+\tilde Z_F\kappa\frac{(q^2)^2}{
1+\delta q^2}\ee
where
\be\label{6.39}
\delta=\frac{\tilde Z_F\kappa}{1-G_A^{(1)}}\ee
is determined by the requirement that for large $q^2$ one expects
$G_A(q)=q^2$ independent of $k$. (This holds up to neglected
logarithmic corrections.) Assuming that for $k\stackrel{\scriptstyle <}
{\sim}
k_{np}$ the term $\bar m^2_A+G_A^{(1)}q^2$ is small as compared to
the $(q^2)^2$ term the approximate form of the propagator
\be\label{6.40}
G_A^{-1}(q)\approx\frac{1}{\tilde Z_F\kappa}
\frac{1}{(q^2)^2}+\frac{1}{q^2}\ee
would be close to the one corresponding to a confining potential.

It is obviously difficult to find an analytical answer to all
these questions and it seems preferable to solve the flow
equation (\ref{6.9}) numerically. A numerical investigation
has been performed by B. Bergerhoff and the
author \cite{BerW}. We show here only a few first results. In fig. 1
we plot the scale dependence of the wave function
renormalization $\tilde Z_F$ as defined by
\be\label{4.41}
\tilde Z_F=\frac{\partial}{\partial q^2}G_A(q)|_{q^2=k^2+2\Lambda
^2_{QCD}}\ee
with $\Lambda_{QCD}$ the two-loop confinement scale. This definition
implements the idea that $\tilde Z_F$ should be taken
constant for very small $k$ (cf. eq. (4.31)) in a smooth way.
One observes the decrease of $\tilde Z_F$ as $k$ is lowered
corresponding to eq. (\ref{6.28}). We have started the running
in the perturbative region at $k=40$ GeV with $G_A(q)_{|k=40\ {\rm
GeV}}$ given by the one-loop perturbative expression (containing
the infrared cutoff). The latter was normalized by $\partial G_A/
\partial q^2|_{q^2=0}=1$, which explains the starting value
of $\tilde Z_F$ (40 GeV) somewhat larger than one for the
definition (\ref{4.41}). In order to
concentrate on the deviation
of $G_A(q)$  from the linear dependence on $q^2$ we
introduce the quantity
\be\label{4.43}
\chi(q)=\frac{\partial}{\partial \ln q^2}\ \ln\left(
\frac{G_A(q)-G_A(0)}{\tilde Z_Fq^2}\right)\ee
Here the mass term $G_A(0)$ is subtracted from
the inverse propagator and the leading perturbative $q^2$-dependence
is divided out. In the classical approximation the expression in
the bracket equals one, and a nonvanishing value of $\chi(q)$
is entirely due to quantum fluctuations. Since a computation
of $\chi(q)$ involves a numerical derivative of a small
difference, one needs a numerical solution of the flow equation
for $G_A(q)$ with a relatively high precision. We observe that
$\chi(q)$ plays the role of a momentum-dependent anomalous
dimension. Within renormalization-group improved
perturbation theory one expects for
$k=0$ (compare eq. (4.26))
\be\label{4.44}
\chi(q)=\frac{13}{6}N_c\frac{g^2(q)}{16\pi^2}\ee
where $g^2(\mu)$ is the running gauge coupling at the
scale $\mu$. In figs. 2 and 3 we compare the numerical determination
of $\chi(q)$ with the renormalization-group improved
one-loop perturbative result (4.44), with
$g$ the two-loop running gauge coupling.
Up to a scaling factor of about 10 \% the two curves
asymptotically coincide for large $q^2$ as $k$ goes to zero.
This is not a trivial result since no assumption of this type
enters the flow equation (\ref{6.9}). We note that a propagator
$\sim q^{-4}$ for small $q^2$ corresponds to $\lim_{k\to0}
\lim_{q^2\to0}\chi(q)=1$. If a $q^{-4}$ behaviour extends
effectively over a certain momentum range $\chi(q)$ should
develop a plateau at one in this range.
As $k$ is lowered we see in fig. 3
a sizeable increase of $\chi(q)$ for small $q^2$.
We also observe a tendency of an extension and flattening
of the maximum somewhat below one. This tendency should
stabilize as $k$ goes to zero. We have stopped the running
at $k=400$ MeV since
for small $k^2$ and $q^2$ the approximations
leading to (4.9) become doubtful. At least part of the momentum
dependence of the gluon vertices according to (A.55)
should presumably be included.

\section{Flow equations for light quarks}
\setcounter{equation}{0}

The formalism for integrating out the gluon degrees of freedom needs not
to be restricted to the heavy quark approximation.
We want to derive in this section the general flow equation
for the quark two-point and four-point function corresponding
to (2.33). For simplicity we mainly consider $N_F$ massless
quarks -- the inclusion of mass terms is straightforward --
and we work in the gauge with $\alpha=0$. At some appropriate
short distance scale we start with the ``classical action'' (cf.
(3.7))
\bear\label{7.1}
&&\Gamma_k[\psi]=Z_\psi\int\frac{d^4q}{(2\pi)^4}\bar\psi(q)\gamma^\mu
q_\mu\psi(q)\ -\frac{1}{2}Z_\psi^2g^2_k\cdot\\
&&\int\frac{d^4p_1}{(2\pi)^4}...\frac{d^4p_4}{(2\pi)^4}(2\pi)^4
\delta(p_1+p_2-p_3-p_4)P^{-1}(p_1-p_3)
\left\{{\cal M}+\frac{1}{(p_1-p_3)^2}{\cal N}\right\}\nonumber\ear
The flow equation then permits to study how $\Gamma_k$ changes its
form as $k$ is lowered. In particular one is interested in pole-like
structures in the quark four-point function which would indicate the
formation of meson bound states \cite{Ell3},\cite{Ell}.

In order to establish the flow equation we have to collect various
pieces which have been discussed in the previous sections. We begin
with the contribution from the gluon and ghost fluctuations $\gamma_A
-\epsilon+\gamma_c$ which only contribute to the four-quark
interaction $\Gamma_{k,4}[\psi]$.
The direct contributions from gluon and ghost loops
read
\be\label{7.2}
\gamma_A-\epsilon=\frac{1}{2}\int\frac{d^4q}{(2\pi)^4}\left(A_k^{(0)}
(q)\right)^z_\mu\left(A^{(0)}_k(-q)\right)^\nu_z
\left\lbrace\frac{\partial}{\partial t}\hat G_A(q)\delta^\mu_\nu
+\frac{\partial}{\partial t}\hat H_A(q)q^\mu q_\nu\right\rbrace\ee
The functions $\frac{\partial}{\partial t}\hat G_A(q)$ and $\frac{\partial}
{\partial t}\hat H_A(q)$ have been computed in the last section
(cf. (\ref{6.1}), (\ref{6.2}), or (\ref{6.9})), where
we have indicated by a hat that only gluon and ghost contributons
should be included here, in analogy  to the heavy quark approximation.
The classical solution $A^{(0)}_k$ is given by eqs. (\ref{5.4})
and (\ref{5.6}), where $G_A$ and $H_A$ characterize now the gluon
propagator without reference to the heavy quark approximation. We consider
again the limit $\alpha\to0$ where $S^\nu_\mu(q)=(G_A+R_k)^{-1}(\delta
^\nu_\mu-q^\nu q_\mu/q^2)$. Including also the contribution $\gamma_c$
which is similar in structure we obtain
\bear\label{7.3}
\gamma_A-\epsilon+\gamma_c&=&-\frac{1}{2}\int\frac{d^4p_1}{(2\pi)^4}...
\frac{d^4p_4}{(2\pi)^4}(2\pi)^4\delta(p_1+p_2-p_3-p_4)\cdot\nonumber\\
&&f_1(p_1,p_2,p_3,p_4)({\cal M}+\frac{1}{(p_1-p_3)^2}
{\cal N})\ear
with
\bear\label{7.4}
&&f_1(p_1,p_2,p_3,p_4)=-G_\psi(-p_1,p_1-p_3)G_\psi(p_4,p_2-p_4)
\cdot\nonumber\\
&&(G_A(p_1-p_3)+R_k(p_1-p_3))^{-2}\left(\frac{\partial}
{\partial t}\hat G_A
(p_1-p_3)+\frac{\partial}{\partial t}R_k(p_1-p_3)\right)\ear
(In the heavy quark approximation there is no difference between
$\hat G_A$ and $G_A$. Also $G_\psi$ is independent of $k$ such that
$f_1=\frac{\partial}{\partial t}F_1$ (\ref{5.8}).)
The function $f_1$ involves the quark gluon vertex $G_\psi$ and
the gluon propagator $G_A$ which also enters in the determination
of $\frac{\partial}{\partial t}\hat G_A$.

Up to this point the only approximations involve the three-
and four-gluon
vertices entering $\frac{\partial}{\partial t}\hat G_A$
and the ghost sector. They  have
been discussed in sect. 4. We may in addition also
truncate the quark gluon
vertex and use for light quarks the ansatz
\be\label{7.5}
G_\psi(p,q)=\tilde Z_F^{1/2}Z_\psi g_k\ee
The running of the renormalized gauge coupling $g_k$ is now determined
by the $\beta$-function including quark contributions and $\tilde Z_F$
may be identified with $G_A^{(1)}$ for $k>k_{np}$ (cf. sect. 4).
the lowest order truncation
for $G_A(q)$ would be
\be\label{7.6}
G_A(q)=\tilde Z_Fq^2\ee
such that $P_A(q)$ is replaced by $P(q)$ in the equation (\ref{6.9})
for $\frac{\partial}{\partial t}\hat G_A$. The detailed discussion
of the last section shows, however, that this approximation becomes
invalid for $k$ in the vicinity of the confinement scale.
There one should
rather use a truncation of the form (\ref{6.38})
or similar. The corresponding
flow equations for $\bar m_A^2, G_A^{(1)}$ and $\kappa$ include
now additional contributions from quark fluctuations and have to be
computed from the evolution equation for $\frac{\partial}{\partial t}
G_A(q)$ in the formulation where both quark and gluon degrees of
freedom are present. Fortunately, the formation of meson-bound states
occurs typically at a scale substantially higher than the confinement
scale. This gives the hope that important features
of meson physics can already be extracted using
the truncation (\ref{7.6}) on the r.h.s. of the flow equation
and do not need a very detailed understanding of gluon condensation
phenomena.

With the truncation (\ref{7.5}), (\ref{7.6}) the function $f_1$
(\ref{7.4}) only depends on the Mandelstam variable $t=(p_1-p_3)^2$
\bear\label{7.7}
f_1&=&-Z^2_\psi g^2_kP^{-2}(p_1-p_3)\nonumber\\
&&\left\lbrace\frac{\partial}{\partial t}
P(p_1-p_3)
-\tilde\eta_F\left(P(p_1-p_3)-(p_1-p_3)^2\right)
+N_cg^2_k{\cal G}
(p_1-p_3)\right\rbrace\ear
where
\be\label{7.8}
\frac{\partial}{\partial t}
\hat G_A(q)=N_cg^2_k\tilde Z_F{\cal G}(q)\ee
is given by the r.h.s. of (\ref{6.9}) with $P_A(q)=P(q)$ and
$\tilde\partial_tP_A(q)=\frac{\partial}{\partial t}P(q)-\tilde\eta_F(P(q)
-q^2)$.
The first two contributions are proportional $\tilde\partial_tP^{-1}$
and simply reflect the change of the infrared cutoff contained
in $P^{-1}$ in the ``classical action '' (\ref{7.1}). Only the last
term $\sim {\cal G}$ describes how additional quantum fluctuations
are included as $k$ is lowered - in this case the gluon and ghost
contributions to the gluon propagator. If we omit the contribution
$\sim\tilde\eta_F$ in $\tilde\partial_tP_A(q)$ the funtion ${\cal G}(q)$
reads explicitly
\bear\label{7.9}
&&{\cal G}(q)=\int\frac{d^4q'}{(2\pi)^4}\frac{\partial}{\partial t}
\Biggl\lbrace\frac{9}{4}P^{-1}(q')-\frac{1}{6}P^{-1}(q')P^{-1}(q+q')
\cdot\nonumber\\
&&\left[13q^2-14(qq')+10{q'}^2-10\frac{(qq')^2}{q^2}-22
\frac{(qq')^2}{{q'}^2}
\right.\nonumber\\
&&\left.-4\frac{(qq')^3}{q^2{q'}^2}+\frac{q^2}{{q'}^2}
\frac{q^2{q'}^2-(qq')^2}{(q+q')^2}\right]\Biggr\rbrace\ear
We observe that even for our simple truncations the function
${\cal G}(q)$
has a complicated momentum dependence. A solution of the flow
equation for the four-quark interaction
will go far beyond the effects of a running gauge
coupling in the one-loop approximation.

The term $\gamma_{A\psi}$ contributes to the flow equation for the
two-point and the four-point function. The contribution to the two-point
function can be extracted from (2.33) using the classical
solution (\ref{5.4}). With $\alpha=0$ and $G_\psi(p,q)=Z_\psi\tilde Z
_F^{1/2}g_k$ the lowest order classical solution reads
\bear\label{7.10}
&&(A_k^{(0)}(q))^\mu_z=-(G_A(q)+R_k(q))^{-1}Z_\psi\tilde Z_F^{1/2}g_k
\nonumber\\
&&\int\frac{d^4p}{(2\pi)^4}\bar\psi^i_a(p)\left(\gamma^\mu-
\frac{q^\mu \slq}{q^2}\right)(T_z)_i^{\ j}\psi^a_j(p+q)\ear
and the next to leading contribution $A^{(1)}_k$ obtains as
\bear\label{7.11}
&&\left(A_k^{(1)}(q)\right)^\mu_z=-ig^3_k\tilde Z_F^{-1/2}Z_\psi^2f_z^{
\ yw}P^{-1}_A(q)\nonumber\\
&&\int\frac{d^4p_1}{(2\pi)^4}...\frac{d^4p_4}{(2\pi)^4}(2\pi)^4
\delta(q-p_1-p_2+p_3+p_4)
P^{-1}_A(p_1-p_3)P^{-1}_A(p_2-p_4)\cdot\nonumber\\
&&\Biggl[\frac{1}{2}(p_1^\mu-p_3^\mu-p_2^\mu+p_4^\mu)
\Biggl(\{\bar\psi(-p_1)\gamma_\nu T_y\psi(-p_3)\}
\{\bar\psi(p_4)\gamma^\nu T_w\psi(p_2)\}\nonumber\\
&&-\frac{(p_1-p_3)_\nu(p_2-p_4)^\nu}
{(p_1-p_3)^2(p_2-p_4)^2}\cdot
\{\bar\psi(-p_1)(\slp_1-\slp_3)T_y\psi(-p_3)\}\{\bar\psi(p_4)
(\slp_2-\slp_4)
T_w\psi(p_2)\}\Biggr)\nonumber\\
&&-\frac{q^\mu}{(p_2-p_4)^2}\{\bar\psi(-p_1)(\slp_2-\slp_4)T_y\psi(-p_3)\}
\{\bar
\psi(p_4)(\slp_2-\slp_4)T_w\psi(p_2)\}\nonumber\\
&&+2\{\bar\psi(-p_1)\left(\slp_2-\slp_4-\frac{(p_2-p_4)_\nu(p_1-p_3)^\nu}
{(p_1-p_3)^2}(\slp_1-\slp_3)\right)T_y\psi(-p_3)\}\nonumber\\
&&\{\bar\psi(p_4)\gamma^\mu T_w\psi(p_2)\}\Biggr]\ear
With the ansatz of a flavour diagonal kinetic term and mass term
\be\label{7.12}\Gamma_k^{(2)}[0]=Z_\psi(c_a(q)\slq+m_a(q)\gamma^5)
\delta_b^a
\delta^i_j(2\pi)^4\delta(q-q')\ee
one finds the following contribution to the flow equation
for the two-point
function
\bear\label{7.13}
&&\gamma_{A\psi}^{(2)}=\frac{N_c^2-1}{2N_c}Z_\psi g^2_k\int
\frac{d^4q}{(2\pi)^4}
\frac{d^4p}{(2\pi)^4}(q+p)^2\left(\frac{\partial}{\partial t}r_k
(q+p)-\tilde\eta_Fr_k(q+p)\right)\nonumber\\
&&P^{-2}_A(q+p)\sum_a\Bigl\{\bar\psi_a(p)[(c_a(q)+r_k^{(\psi)}
(q))^2q^2+m^2_a(q)]^{-1}\nonumber\\
&&[(c_a(q)+r_k^{(\psi)}(q))(\slq+2\frac{q^2+(qp)}{(q+p)^2}(\slq+\slp))
-3m_a(q)\gamma^5]\psi_a(p)\Bigr\}\ear
The contribution to the four-point function $\gamma^{(4)}_{A\psi}$
can be obtained similarly using (5.11).

For a computation of the purely fermionic contribution $\gamma_\psi$ we
will use the following truncation for the term quartic in the fermionic
fields
\bear\label{7.14}
&&\Gamma_{k,4}^{(\psi)}=-Z^2_\psi\int\frac{d^4p_1}{(2\pi)^4}
...\frac{d^4p_4}{(2\pi)^4}(2\pi)^4\delta(p_1+p_2-p_3-p_4)\nonumber\\
&&\left\{\lambda_\sigma(p_1,p_2,p_3,p_4){\cal M}_\sigma+\lambda_\rho
(p_1,p_2,p_3,p_4){\cal M}_\rho\right.\nonumber\\
&&\left.+\lambda_p(p_1,p_2,p_3,p_4){\cal M}_p+
\lambda_n(p_1,p_2,p_3,p_4)
{\cal N}\right\}\ear

This yields a contribution to the flow
equation for the two-point function
\bear\label{7.15}
\gamma_\psi^{(2)}&=&Z_\psi\int\frac{d^4p}{(2\pi)^4}\sum_a
\biggl\{\bar\psi_a(p)\int
\frac{d^4q}{(2\pi)^4}\left(\frac{\partial}{\partial t}
r_k^{(\psi)}(q)-\eta_\psi r_k^{(\psi)}(q)\right)\nonumber\\
&&\left[\left(c_a(q)+r_k^{(\psi)}(q)\right)^2 q^2+m^2_a(q)\right]^{-2}
\Biggl\{
\left[\left(c_a(q)+r_k^{(\psi)}(q)\right)^2q^2-m^2_a(q)\right]\nonumber\\
&&\Bigl[ 2N_c\lambda_\rho(-q,q,-p,p) \slq-\frac{2}{N_c} \lambda_p
(-q,q,-p,p)\slq\nonumber\\
&&-\frac{N_c^2-1}{N_c}\lambda_n(-q,q,-p,p)\left(\left(p^2-q^2\right)
\slq+2
\left(q^2-(pq)\right) \slp\right)\Bigr]\nonumber\\
&&+m_a(q)\left( c_a(q)+r_k^{(\psi)}(q)\right) q^2\gamma^5
\Bigl[ 8N_c\lambda_\sigma
(-q,q,-p,p)\nonumber\\
&&
-\frac{8}{N_c}\lambda_p(-q,q,-p,p)-\frac{2(N_c^2-1)}{N_c}(q-p)^2
\lambda_n(-q,q,-p,p)\Bigr]\Biggr\} \psi^a(p)\biggr\}\nonumber\\
&&+Z_\psi\int\frac{d^4p}{(2\pi)^4}\frac{d^4q}{(2\pi)^4}
\left\{\bar\psi_b(p)\slq\psi
^b(p)\right\}\left(\frac{\partial}{\partial t}r^{(\psi)}_k(q)-\eta_\psi
r_k^{(\psi)}(q)\right)\nonumber\\
&&\sum_a\frac{(c_a(q)+r_k^{(\psi)}(q))^2q^2-m^2_a(q)}{[(c_a(q)+
r^{(\psi)}_k(q))^2+m^2_a(q)]^2}
\left[2\lambda_\sigma(-q,p,-q,p)\right.\nonumber\\
&&\left.+2\lambda_\rho(-q,p,-q,p)-4\lambda_p(-q,p,-q,p)\right].\ear
For $\gamma_{(\psi)}^{(4)}$ we present here only the case  
$\lambda_\rho=\lambda_p=\lambda_n=0$
in the chiral limit $(m_a=0)$. Omitting all contributions except
$\gamma_\psi^{(4)}$ one finds
\bear\label{7.15a}
&&\frac{\partial}{\partial t}\lambda_\sigma(p_1,p_2,p_3,p_4)=
2\eta_\psi\lambda
_\sigma(p_1,p_2,p_3,p_4)\nonumber\\
&&+8N_c\int\frac{d^4q}{(2\pi)^4}\frac{q_\mu(q-p_1-p_2)^\mu}
{q^2(q-p_1-p_2)^2}
[c(q)+r_k^{(\psi)}(q)]^{-1}\nonumber\\
&&\tilde\partial_t[c(q-p_1-p_2)+r^{(\psi)}_k(q-p_1
-p_2)]^{-1}\nonumber\\
&&\lambda_\sigma(p_1,p_2,q,-q+p_1+p_2)\lambda_\sigma
(q,-q+p_1+p_2,p_3,p_4)
\ear
For the special case $(c(q)+r_k^{(\psi)}(q))^{-1}=\exp\left(-\frac{q^2}
{\Lambda^2}\right)-\exp\left(-\frac{q^2}{k^2}\right),\eta_\psi=0$
this reproduces the flow equation of ref. \cite{Ell}. The r.h.s. of
this equation should now be supplemented by the contributions from
$\gamma_A-\epsilon+\gamma_c+\gamma_{A\psi}$ which have been
discussed before. Furthermore, a better approximation should include
the contributions from $\lambda_\rho$ etc.

Besides this, eq. (\ref{7.15a}) involves the explicit momentum
dependence of the fermion kinetic term, i.e. the function
$c(q)$. The scale dependence of the quark propagator
can be computed from
$\gamma_\psi^{(2)}$ and $\gamma_{A\psi}^{(2)}$. Combining eq. (\ref{7.15})
with (\ref{7.13}) and (\ref{7.12}) yields the flow equation
for the kinetic term
\bear\label{7.16}
&&\frac{\partial}{\partial t}\left( Z_\psi c_a(p)\right)= Z_\psi \int
\frac{ d^4 q}{(2\pi)^4}\Biggl\{\frac{N^2_c-1}{2 N_c} g^2_k
\left(\frac{\partial}{\partial t} r_k (p-q)-\tilde\eta_F r_k
(p-q)\right)\nonumber\\
&&
P_A^{-2}(p-q)\left[\left( c_a(q)+r_k^{(\psi)}(q)\right)^2q^2+
m^2_a(q)\right]^{-1}\nonumber\\
&&\left(c_a(q)+r_k
^{(\psi)}(q)\right) \left( 2q^2-3(pq)-3\frac{(pq)q^2}{p^2}+4
\frac{(pq)^2}{p^2}\right)\nonumber\\
&&-\left(\frac{\partial}{\partial t} r_k^{(\psi)}(q)-\eta_\psi r_k^{(\psi)}
(q)\right)
\left[\left( c_a(q)+r_k^{(\psi)}(q)\right)^2q^2+m_a^2(q)\right]^{-2}
\nonumber\\
&&\left[\left(c_a(q)+r_k^{(\psi)}(q)\right)^2q^2-m^2_a(q)\right]
\Biggl[\left(2N_c\lambda_\rho(-q,q,-p,p)-\frac{2}{N_c}\lambda_p(-q,q,-p,p)
\right)\frac{(pq)}{p^2}\nonumber\\
&&
-\frac{N^2_c-1}{N_c}\lambda_n(-q,q,-p,p)\left(2q^2-(pq)-
\frac{(pq)q^2}{p^2}\right)
\Biggr]\nonumber\\
&&+\frac{(pq)}{p^2}\left(\frac{\partial}{\partial t}r_k^{(\psi)}(q)-
\eta_\psi
r^{(\psi)}_k(q)\right)\sum^{N_F}_{b=1}\frac{(c_b(q)+r_k^{(\psi)}(q))
^2q^2-m_b^2(q)}{[(c_b(q)+r^{(\psi)}_k(q))^2+m_b^2(q)]^2}\cdot\nonumber\\
&&[2\lambda_\sigma(-q,p,-q,p)+2\lambda_\rho(-q,p,-q,p)-4\lambda_p(-q,
p,-q,p)]\Biggr\}
\ear
The right-hand side of this equation involves the gauge coupling $g_k$
and the
effective inverse gauge field propagator $P_A$ as well as the effective
fermionic four-point vertices $\lambda_\sigma,
\lambda_\rho,\lambda_p$ and $\lambda_n$.
It is instructive to study this equation in the ``classical approximation''
for the  four-quark vertices $\lambda_\sigma$, i.e.
\bear\label{7.17}
\lambda_\sigma(p_1,p_2,p_3,p_4)&=&\lambda_\rho(p_1,p_2,p_3,p_4)=\lambda_p
(p_1,p_2,p_3,p_4)\nonumber\\
&=&\frac{1}{4} g^2_k\left(P_A^{-1}(p_1-p_3)+P_A^{-1}(p_2-p_4)\right)\\
\lambda_n(p_1,p_2,p_3,p_4)&=&\frac{1}{4}g^2_k\left(\frac{1}{(p_1-p_3)^2
P_A(p_1-p_3)}+\frac{1}{(p_2-p_4)^2P_A(p_2-p_4)}\right)\nonumber
\ear
Using
\bear\label{7.18}
\tilde\partial_t P_A^{-1}(q)&=&-q^2 P_A^{-2}(q)\left(\frac{\partial}
{\partial t}r_k(q)-\tilde\eta_F r_k(q)\right)\nonumber\\
\tilde\partial_t\frac{c_a(q)+r_k^{(\psi)}(q)}{(c_a(q)+r_k^{(\psi)}
(q))^2q^2+m^2_a(q)}&=&
-\frac{(c_a(q)+r_k^{(\psi)}(q))^2q^2-m_a^2(q)}{((c_a(q)+r_k^{(\psi)}
(q))^2q^2+m_a^2(q))^2}\cdot\nonumber\\
&&\left(\frac{\partial}{\partial t}r_k^{(\psi)}(q)-\eta_\psi r_k^{(\psi)}
(q)\right)\ear
one obtains
\bear\label{7.19}
\frac{\partial}{\partial t}(Z_\psi c_a(p))&=&Z_\psi\frac{N_c^2-1}{2N_c}
g^2_k\int \frac{d^4 q}{(2\pi)^4}\left(\frac{2(pq)}{p^2}-
\frac{2q^2p^2-(pq)(p^2+q^2)}{p^2(p-q)^2}
\right)\cdot\nonumber\\
&&\tilde\partial_t\left( P_A^{-1}(p-q)\frac{c_a(q)+r_k^{(\psi)}
(q)}{(c_a(q)+r_k^{(\psi)}
(q))^2q^2+m^2_a(q)}\right)
\ear
We observe that this expression corresponds to the formal
$\tilde \partial_t$
derivative of the standard one-loop correction to the fermion kinetic term
in presence of an infrared cutoff in the propagator.

The fermionic wave function renormalization $Z_\psi$
can be defined by $c_a(p_0)=1$
for suitable $p_0$ and $a$. We will choose here $p_0=0$ and use $c_a$
corresponding to a light quark. Defining the anomalous dimension
\be\label{7.20}
\eta_\psi=-\frac{\partial}{\partial t}\ln Z_\psi\ee
one obtains for $m_a=0$
\be\label{7.21}
\eta_\psi=\frac{3}{4}\frac{N_c^2-1}{N_c}g^2_k\int\frac{d^4q}{(2\pi)^4}
\frac{1}{q^2}\tilde\partial_t\left\{(c(q)+r_k^{(\psi)}(q))^{-1}
P^{-1}_A(q)(1-q^2\dot P_A(q)P^{-1}_A(q))\right\}\ee
where $\dot P_A=\partial P_A/\partial q^2$.
In the perturbative limit $c(q)=1,\ P_A(q)=P(q),\tilde\partial_tP_A(q)=
\frac{\partial}{\partial t}P(q)$
this yields a vanishing fermionic dimension
\be\label{7.22}
\eta_\psi=0\ee

We also may extract the anomalous mass dimension by expanding
eqs. (\ref{7.13}) and (\ref{7.16}) in linear order in $m_a$
\be\label{7.23}
\frac{\partial}{\partial t}m_a(0)=\omega_mm_a(0)\ee
One obtains
\bear\label{7.24}
&&\omega_m=\int\frac{d^4q}{(2\pi)^4}\frac{m(q)}{m(0)}\Biggl\{\frac{3}{2}
\frac{N_c^2-1}{N_c}g^2_k\frac{1}{(c(q)+r_k^{(\psi)}(q))^2q^2}\tilde\partial_t
P_A^{-1}\\
&&+\tilde\partial_t\frac{1}{(c(q)+r^{(\psi)}_k(q))^2q^2}[4N_c\lambda_\sigma
(-q,q,0,0)-\frac{4}{N_c}\lambda_p(-q,q,0,0)\nonumber\\
&&-\frac{N_c^2-1}{N_c}q^2\lambda_n
(-q,q,0,0)]\Biggr\}\nonumber\ear
which reduces in the ``classical approximation''
$\lambda_\sigma(-q,q,0,0)=\lambda_p(-q,q,0,0)=q^2_k\lambda_n(-q,q,0,0)=
\frac{1}{2}g^2P_A^{-1}(q)$ and for $m_a(q)=m_a(0)$ to
\be\label{7.25}
\omega_m=\frac{3}{2}\frac{N_c^2-1}{N_c}g^2_k\int\frac{d^4q}{(2\pi)^4}\tilde
\partial_t\frac{1}{(c(q)+r_k^{(\psi)}(q))^2q^2P_A(q)}\ee
In lowest order perturbation theory where
$\tilde\partial_t[(c(q)+r_k^{(\psi)}(q))^2q^2P_A(q)]^{-1}=\frac{\partial}
{\partial t}[(1+r_k^{(\psi)}(q))^2q^2P(q)]^{-1}$ one recovers the standard
perturbative result
\be\label{7.26}
\omega_m=-\frac{3}{16\pi^2}\frac{N_c^2-1}{N_c}g^2_k\ee
provided $\lim_{q^2\to0}q^2/(1+r_k^{(\psi)}(q))^2=0$.

To summarize this section, we have explicitly derived a nonperturbative
flow equation for the scale dependence of the quark propagator (\ref{7.16}).
We have also given a simplified non-perturbative approximation (\ref{7.19})
to this evolution equation and checked explicitly the consistency with
perturbation theory in the region where the gauge coupling $g_k$ is small.
In the approximation (\ref{7.19}) the flow equation ressembles a
differential
form of the Schwinger-Dyson equation for the full momentum-dependent
quark propagator. In contrast to the usual Schwinger-Dyson equation we
notice that only a small range in momenta $q^2\approx k^2$ contributes
to the momentum integral on the r.h.s. of eq. (\ref{7.19}). This is
therefore much easier to control. The actual complicated scale and momentum
dependence of $Z_\psi c(q)$, which reflects that the physical picture
varies from high to low scales, arises then as a property of the solution
of the differential equation (\ref{7.19}) rather than through the form of
the equation itself. This solution will provide the anomalous dimension
$\eta_\psi$  and the function $c(q)$ needed for the flow equation of the
quark four-point function. We have explicitly computed the contributions
$\gamma_A-\epsilon+\gamma_c$ to the non-perturbative flow equation for
the quark four-point function (\ref{7.3}), (\ref{7.7}), (\ref{7.9}).
It is straightforward to extract the contributions from $\gamma_{A\psi}$
and $\gamma_\psi$ from (\ref{7.11}) and (5.16).
(The latter may be generalized to include $\lambda_\rho$ etc..)
The result gives
the non-perturbative evolution equation for the momentum-dependent
four-quark coupling.

\section{Conclusions and discussion}
\setcounter{equation}{0}

In this paper we have developed a formalism for integrating out the gluon
fields in order to obtain an effective action for the quarks. This is not
done at once since such an approach would lead to complicated nonlocalities
and a reliable direct computation seems almost impossible.
Instead, we account for the gluon contributions to an exact
nonperturbative flow equation. At every scale $k$ this needs only
information about the ``classical solution'' for the
gauge field in presence
of fermions and about quadratic gauge field fluctuations around this
solution with momenta $q^2\approx k^2$. The flow equation describes
the scale dependence of an effective average action $\Gamma_k[\psi]$
which only involves the quark fields. In the present work we have
concentrated on analytical work whereas the numerical exploitation
of our formulae is left to a separate investigaton \cite{BerW}.

Our first main result is the evolution equation (6.9) for the gluon
propagator in the heavy quark limit. As described in sect. 3, its
quantitative solution is connected with the heavy quark
potential. The second result concerns the effective action for
light quarks. The evolution equation for the two- and four-point
functions can be extracted from sect. 5. The solution for the
four-point function is expected to develop pole-like structures
connected to
mesons. This can be treated with the composite-field methods developed
in ref. \cite{Ell}, such that one can finally make a transition to
an effective theory for mesons describing the low momentum
behaviour of QCD. The main difference of the present formalism as
compared to ref. \cite{Ell} concerns the treatment of the gluons: While
ref. \cite{Ell} accounts for the effects of gluons only by a
phenomenologically
motivated four-quark interaction, no such term is introduced here by
hand. At short distances we simply start with the QCD action
for quarks and
gluons. The effective four-quark interaction should arise as a
property of the solution of the flow equation. For this purpose it is
crucial that the gluon contributions to the evolution equation are
properly taken into account.

In order to obtain the $k$-dependent classical solution for the gluon
$A_\mu$ as a functional of the quark fields $\psi$ we need knowledge
about the effective action $\Gamma_k[\psi,A]$ for both quarks and
gluons. The same holds true for the quadratic fluctuations of the gluon
field around this solution. More concretely, the exact evolution
equations for the fermionic two- and four-point functions involve the
gluon propagator for $\psi=0$, the $\bar\psi\psi A, (\bar\psi\psi)^2A,
(\bar\psi\psi)A^2$, $(\bar\psi\psi)^2A^2$ and $\bar\psi\psi A^3$
vertices as well
as the gluonic vertices $A^3$ and $A^4$ in $\Gamma_k[\psi,A]$.
Obviously, these quantities can be computed only approximately
and truncations are needed. We propose to use the nonperturbative
evolution equation for $\Gamma_k[\psi,A]$ in order to compute the inverse
propagator $(\sim A^2)$
and at least one vertex $(\sim A^3)$. (Other vertices can then be related
to the $A^3$ vertex). Now the reader may ask why we do
not work entirely in the framework of the evolution equation for
$\Gamma_k[\psi,A]$, extracting $\Gamma[\psi]$ only at
the end of the evolution for $k=0$. Indeed, we have recovered
the perturbative $\beta$-functions as limiting cases of our
nonperturbative flow equations for small gauge couplings, and these
perturbative $\beta$-functions are certainly easier obtained in
the framework of the evolution equation for $\Gamma_k[\psi,A]$.
Also, the heavy quark potential only involves a computation of the
gluon propagator encoded in $\Gamma_k[\psi,A]$. The main virtue of our
approach concerns the nonperturbative aspects of the flow equations
with light quarks: We want to obtain a quantitatively reliable flow
equation for the full momentum dependence of the quark four-point
function. This encodes the formation of meson-bound states as pole-like
structures in the $s$-channel. In turn, this requires a control of
the momentum dependence on the r.h.s. of the flow equation. Using a
flow equation for $\Gamma_k[\psi,A]$ we would need a computation
of the momentum dependence of the effective vertices $\bar\psi\psi A,(\bar
\psi\psi)^2A^2,(\bar\psi\psi)A^2,(\bar\psi\psi)^2A^2,\bar\psi
\psi A^3$, $A^3$ and $A^4$
in addition to the momentum dependence of the gluon and quark propagator
and the $(\bar\psi\psi)^2$ vertex in $\Gamma_k[\psi,A]$. Using the
effective
average action for quarks $\Gamma_k[\psi]$ a large part of this momentum
dependence is encoded in the two- and four-point functions in
$\Gamma_k[\psi]$. (Note that the effective $(\bar\psi\psi)^2$ vertex in
$\Gamma_k[\psi]$ is different from the corresponding one in
$\Gamma_k[\psi,A]$ since effects of mixed quark-gluon vertices are
included through the classical solution for $A$.) One may therefore
hope that the momentum dependence of the propagator and four-quark
interaction in $\Gamma_k[\psi]$ includes the dominant effects, whereas
a less precise estimate is sufficient for the vertices appearing
in the contributions from the gluon fluctuations around the classical
solution. In the truncation used in this paper we neglect for a computation
of the classical solution and the gluon fluctuations the vertices $\bar
\psi\psi A^2,(\bar\psi\psi)^2A,(\bar\psi\psi)^2A^2$ and $\bar\psi\psi A^3$
and we
describe the three vertices $\bar\psi\psi A,A^3$ and $A^4$ by one common
scale-dependent, but momentum-independent, coupling constant $\tilde g
(k)$. Clearly, establishing the flow equations for the momentum dependence
of the $\bar\psi\psi A,A^3$ and $A^4$ vertices and using the appropriate
solution on the r.h.s. of the flow equations should be one of the next
steps in our approach.

One may suspect that even the non-perturbative treatment of this
paper breaks down for scales $k$ of the order or below the confinement
scale.
Fortunately the formation of meson-bound states is expected at a
scale $k_\varphi$ considerably higher than the confinement scale.
For the pseudoscalar mesons a first computation indicates
$k_\varphi\approx$
650 MeV \cite{Ell}. In view of this one may hope that an
understanding of the formation of mesons does not necessitate
a very detailed understanding of the physics near the confinement scale.
There are still several ``hopes'' and ``expectations''.
A quantitative computation of the quark condensate $<\bar\psi\psi>$
and the pion decay constant $f_\pi$ along similar lines as in
ref. \cite{Ell}, but using the flow equations proposed in the present
paper should decide whether we are on a reasonable track for an
analytical understanding of QCD.
\bigskip

\noindent {\bf Note added}: Results for the form of the heavy
quark potential have recently been obtained using flow equations
\cite{Eneu} and are closely related to sect. 4 of the present work.

\section*{Appendix}
\renewcommand{\theequation}{A.\arabic{equation}}
\setcounter{equation}{0}
For the derivation of general identities it is convenient to work
with a generalized field $\chi$
\bear\label{A.1}
&&\chi=(a^z_\mu, -c^z,\bar c^z,-\psi_m',\bar\psi_m',\varphi_a',
\varphi_a^{*'})\nonumber\\
&&\bar\chi=(a^{*z}_\mu, \bar c^z,c^z,\bar\psi_m',\psi_m',\varphi_{a}
^{*'},\varphi_a')\ear
It is composed of real gauge fields $a^z_\mu={\cal A}^z_\mu-\bar A^z_\mu$,
ghosts $c^z$ and antighosts $\bar c^z$ as well as complex spinors $\psi_m$
and complex scalars $\varphi_a$ in some representations of the gauge
group. We use here a notation which can be employed both in coordinate
space $a(x)=a^*(x)$ and in momentum space where $a^*(p)=a(-p)$ and
$\delta\chi^{\tilde\alpha}(p)/\delta\chi^{\tilde\beta}(p')=
(2\pi)^d\delta(p-p')\delta^{\tilde\alpha\tilde\beta}$.
For real scalars or Majorana spinors one should omit the doubling of
arguments in $\chi$ and impose ${\varphi^*}'(p)=\varphi'(-p)$
or similar for Majorana spinors. In our notation the quadratic
part of the action reads
\be\label{A.2}
S_2=\frac{1}{2}\int\frac{d^dp}{(2\pi)^d}\frac{d^dp'}{(2\pi)^d}
S^{(2)}_{\tilde
\alpha\tilde\beta}(p,p')\chi_{\tilde\beta}(p')\bar
\chi_{\tilde\alpha}(p)\equiv\frac{1}{2}S^{(2)}_{\alpha\beta}
\chi_\beta\bar\chi_\alpha\ee
\be\label{A.3}
S^{(2)}_{\tilde\alpha\tilde\beta}(p,p')=\frac{\delta^2S}
{\delta\bar\chi_
{\tilde\alpha}(p)\delta\chi_{\tilde\beta}(p')}\ee
In the last part of (A.2) we have combined internal indices
$\tilde\alpha$ and momentum labels $p$ to a collective index
$\alpha$. The infrared cutoff is introduced as a quadratic block
diagonal piece in the action
\bear\label{A.4}
&&\Delta_kS=\frac{1}{2}{\R}_{k\alpha\beta}\chi_\beta
\bar\chi_\alpha\nonumber\\
&&{\R}_k=diag({\R}_k^{(A)},{\R}_k^{(c)},{\R}_k^{(c)^*},
{\R}_k^{(\psi)},{\R}_k^{(\psi)*},{\R}_k^{(\varphi)},
{\R}_k^{(\varphi)^*})={\R}_k^\dagger
\ear
with ${\R}^{(A)^*}_{yz}(p,p')={\R}^{(A)}_{yz}(-p,-p')$.
We also introduce sources
\bear\label{A.5}
&&J=(K,\zeta,\bar\zeta,\eta,\bar\eta,j,j^*)\nonumber\\
&&\bar J=(K^*,-\bar\zeta,\zeta,-\bar\eta,\eta,j^*,j)\ear
and define
\bear\label{A.6}
e^{W_k[J]}&=&\int{\cal D}\chi e^{-S}\nonumber\\
S&=&S_0+S_{\gf}+S_{\gh}+\Delta_kS+S_{\rm source}\ear
Here $S_0$ is the classical gauge-invariant action, $S_{\gf}$ and
$S_{\gh}$ are gauge-fixing and ghost terms and
\be\label{A.7}
S_{\rm source}=-\bar J_\alpha \chi_\alpha=-J_\alpha\bar\chi_\alpha\ee
For the Legendre transform
$\tilde \Gamma_k=\bar J_\alpha\sigma_\alpha-W_k$
the following identities hold
\be\label{A.8}
\sigma_\alpha=\frac{\partial W_k}{\partial \bar J_\alpha},
\quad \bar\sigma_\alpha=\frac{\partial W_k}{\partial J_\alpha}\ee
\be\label{A.9}
\frac{\partial\tilde\Gamma_k}{\partial\sigma_\alpha}=M_{\alpha\beta}\bar
J_\beta,\quad
\frac{\partial\tilde\Gamma_k}{\partial\bar\sigma_\alpha}
=M_{\alpha\beta}J_\beta,\ee
\be\label{A.10}
\frac{\partial^2W_k}{\partial\bar J_\alpha\partial J_\beta}
\quad\frac{\partial^2\tilde\Gamma_k}{\partial\bar\sigma_\beta
\partial\sigma_\gamma}=
\frac{\partial^2W_k}{\partial J_\alpha\partial \bar J_\beta}
\quad\frac{\partial^2\tilde\Gamma_k}{\partial\sigma_\beta
\partial\bar\sigma_\gamma}=M_{\alpha\gamma}\ee
The appearance of the matrix $M=diag(1,-1,-1,-1,-1,1,1)$
reflects the anticommuting properties of the Grassmann variables
$\xi,\bar\xi,\psi$ and $\bar\psi$ in a notation where
\bear\label{A.11}
\sigma&=&(\bar a,-\xi,\bar\xi,-\psi,\bar\psi,\varphi,\varphi^*)
\nonumber\\
\bar\sigma&=&(\bar a^*,\bar\xi,\xi,\bar\psi,\psi,\varphi^*,\varphi)
\ear
and $\bar a=A-\bar A$.
Taking a derivative of (A.6) with respect to $t=\ln k$ and noting
that only $\Delta_kS$ depends on $k$, one finds \cite{4E} the
flow equation for $\Gamma_k=\tilde\Gamma_k-\frac{1}{2}{\cal R}
_{\alpha\beta}\sigma_\beta\bar\sigma_\alpha$
\be\label{A.12}
\partial_t\Gamma_k=\frac{1}{2}S{\Tr}\left\{\partial_t{\cal R}_k
\left(\Gamma_k^{(2)}+{\cal R}_k\right)^{-1}\right\}\ee
with $S{\rm Tr} A={\Tr}MA,\ S{\Tr}AB=S{\Tr}BA$, and
${\rm Tr}=\int\frac{d^dp}{(2\pi)^d}\sum_{\tilde\alpha}$.
The inverse propagator $\Gamma^{(2)}=\tilde\Gamma^{(2)}-{\cal R}_k$
is given by
\be\label{A.13}
\left(\Gamma_k^{(2)}\right)_{\alpha\beta}=\frac{\partial^2\Gamma_k}
{\partial\bar\sigma_\alpha\partial\sigma_\beta}\ee

We next turn to the anomalous Ward-Takahashi or Slavnov-Taylor
identities for which we follow closely the treatment of ref. \cite{Ell2}.
For a general gauge-fixing $G^z(a)$ linear in $a^z_\mu$ (and
possibly depending in addition on the background field $\bar A
^z_\mu$)
\bear\label{A.14}
&&S_{\gf}=\frac{1}{2\alpha}\int d^dxG^zG^z\nonumber\\
&&S_{\gh}=-\int d^dx\bar c^z\frac{\partial G^z}
{\partial a^y_\mu}(D_\mu(a+\bar A))^{yw}c^w\ear
the sum $S_0+S_{\gf}+S_{\gh}$ is invariant under the BRS variation $\chi
\to\chi+\delta_{\rm BRS}\chi\epsilon, \bar\chi\to
\bar\chi+\delta_{\rm BRS}\bar\chi\varepsilon$
\bear\label{A.15}
&&\delta_{\rm BRS}\chi=\left(\frac{1}{\bar g}(D_\mu(a+\bar A)c)^z,-  
\frac{1}{2}f^{zyw}c^yc^w,\ -\frac{1}{\alpha\bar g}G^z,\right.
\nonumber\\
&&ic^z(T_z)_{mn}\psi_n',-ic^z(T_z^*)_{mn}\bar\psi_n',
ic^z(T_z)_{ab}\varphi_b',-ic^z(T_z^*)_{ab}\varphi^{*'}_b
\Biggr)\nonumber\\
&&\delta_{\rm BRS}\bar\chi=\left(\frac{1}{\bar g}(D_\mu(a+\bar A)c)^{*z},-  
\frac{1}{\alpha\bar g}G^z,\frac{1}{2}f^{zyw}c^yc^w,\right.
\nonumber\\
&&-ic^z(T_z^*)_{mn}\bar\psi_n',-ic^z(T_z)_{mn}\psi_n',
-ic^z(T_z^*)_{ab}{\varphi_b^*}',ic^z(T_z^*)_{ab}\varphi_b'
\Biggr)
\ear
Here $f^{zyw}$ are the structure constants of the gauge group
and $T_z$ the hermitean
generators in the appropriate representations.
It is useful to introduce external sources $(\bar\beta^z_\mu(p)=\beta_\mu^z(-p))$
\bear\label{A.16}
\beta&=&(\beta_\mu^z,\gamma^z,\bar\gamma^z,
\delta_m^{(\psi)},\bar\delta_m^{(\psi)},\delta^{(\varphi)}_a,
\bar\delta^\varphi_a)\nonumber\\
\bar\beta&=&(\bar\beta_\mu^z,-\bar\gamma^z,\gamma^z,-\bar
\delta_m^{(\psi)},\delta_m^{(\psi)},\bar\delta^{(\varphi)}_a,
\delta^{(\varphi)}_a)\ear
for the BRS variations of $\chi$ or $\bar\chi$ and to define $W_k[J,\bar\beta]$
similar to (\ref{A.6}) by adding in (\ref{A.7}) an additional
source term
\be\label{A.17}
S_{\rm source}^{(\bar\beta)}=-\bar\beta_\alpha(\delta_{\rm BRS}
\chi)_\alpha=-\beta_\alpha(\delta_{\rm BRS}
\bar\chi)_\alpha\ee
The BRS-invariance of the measure implies
\bear\label{A.18}
0&=&<\delta_{\rm BRS}S>=<{\cal R}_{\alpha\beta}\chi_\beta
(\delta_{\rm BRS}\bar\chi)_\alpha>_{|\gamma=0}
-< J_\alpha(\delta_{\rm BRS}\bar\chi)_\alpha>_{|\gamma=0}\nonumber\\
&=&<{\cal R}_{\alpha\beta}(\delta_{\rm BRS}\chi)_\beta
\bar\chi_\gamma M_{\gamma\alpha}>_{|\gamma=0}-<\bar J_\alpha
(\delta_{\rm BRS}\chi)_\alpha>_{|\gamma=0}
\ear
or
\be\label{A.19}
J_\alpha\frac{\partial W_k}{\partial\beta_\alpha}_{|\gamma
=0}={\cal R}_{\alpha\beta}\left(\frac{\partial}{\partial \bar J_\beta}
+\frac{\partial W_k}{\partial \bar J_\beta}\right)
\frac{\partial W_k}{\partial \beta_\alpha}_{|\gamma=0}\ee
Using (\ref{A.8}), (\ref{A.9}) and the identities
\be\label{A.20}
\frac{\partial W_k}{\partial\beta_\alpha}_{|J}=-\frac{\partial
\tilde\Gamma_k}{\partial\beta_\alpha}_{|\bar\sigma}\ee
\be\label{A.21}
\frac{\partial^2W_k}{\partial \bar J_\alpha\partial\beta_\gamma}=-
\frac{\partial^2W_k}{\partial \bar J_\alpha\partial J_\beta}
\frac{\partial^2\tilde\Gamma}{\partial\bar\sigma_\beta
\partial\beta_\gamma}\ee
one arrives at the identity
\bear\label{A.22}
&&\frac{\partial\Gamma_k}{\partial\bar\sigma_\alpha}M_{\alpha\beta}
\frac{\partial\Gamma_k}{\partial\beta_\beta}_{|\gamma=0}=
\frac{\partial\Gamma_k}{\partial\sigma_\alpha}M_{\alpha\beta}
\frac{\partial\Gamma_k}{\partial\bar\beta_\beta}_{|\gamma=0}=S {\rm Tr}
\left\{{\cal R}(\Gamma_k^{(2)}+{\cal R})^{-1}\frac{\partial^2
\Gamma_k}{\partial\bar\sigma\partial\beta}
\right\}_{|\gamma=0}\nonumber\\
&&=M_{\alpha\beta}{\cal R}_{\beta\gamma}(\Gamma^{(2)}+{\cal R})^{-1}
_{\gamma\delta}\frac{\partial^2\Gamma_k}{\partial\bar\sigma_\delta
\partial\beta_\alpha}_{|\gamma=0}\ear

In addition, we note the simple identity
\be\label{A.23}
\frac{\partial\Gamma_k}{\partial\gamma^z}=\frac{1}{\alpha\bar g}
G^z(\bar a)\ee
which allows to eliminate the source $\gamma$. Similarly,
linearity in $\bar c$ yields the field equation for the antighost
\be\label{A.24}
\frac{\partial\Gamma_k}{\partial\bar\xi^z}=\bar g\tilde G_\mu^{zy}
\frac{\partial\Gamma_k}{\partial\bar\beta_\mu^y}\ee
where
\be\label{A.25}
\tilde G^{zy}_\mu=\frac{\partial G^z}{\partial a^y_\mu}=(D_\mu(\bar A))
^{zy}\ee
depends only on the background field $\bar A$. For the last
identity in (\ref{A.25}) we have used the
particular gauge-fixing $G^z=
[D_\mu(\bar A)]^{zy}a^y_\mu$, which will be assumed in the
following. For this gauge we now insert (A.23), (A.24) such that the
Ward identity reads in explicit components in momentum space
\cite{Ell2}
\bear\label{A.26}
&&\int\frac{d^dp}{(2\pi)^d}\left\{\frac{\partial\Gamma'}{
\partial A^z_\mu(p)}\frac{\partial\Gamma'}{
\partial \bar\beta^z_\mu(p)}-\frac{\partial\Gamma'}{
\partial\xi^z(p)}\frac{\partial\Gamma'}{
\partial \bar\gamma^z(p)}\right.\nonumber\\
&&-\frac{\partial\Gamma'}{
\partial \psi_m(p)}\frac{\partial\Gamma'}{
\partial \bar\delta^{(\psi)}_m(p)}-\frac{\partial\Gamma'}{
\partial \bar\psi_m(p)}\frac{\partial\Gamma'}{
\partial\delta^{(\psi)}_m(p)}\nonumber\\
&&\left.+\frac{\partial\Gamma'}{
\partial \varphi_a(p)}\frac{\partial\Gamma'}{
\partial\bar\delta^{(\varphi)}_a(p)}+\frac{\partial\Gamma'}{
\partial\varphi^*_a(p)}\frac{\partial\Gamma'}{
\partial\delta^{(\varphi)}_a(p)}\right\}={\cal A}^{(g)}_{\rm BRS}
+{\cal A}^{(m)}_{\rm BRS}\ear
with
\bear\label{A.27}
&&{\cal A}^{(g)}_{\rm BRS}=\int\frac{d^dp}{(2\pi)^d}\frac{
d^dp'}{(2\pi)^d}
\frac{d^dq}{(2\pi)^d}\left\{{\R}^{(A)\mu\nu}_{zy}
(p,p')(\Gamma^{(2)}+{\cal R})^{-1}_{A^y_\nu(-p')\sigma
_{\tilde\alpha}(q)}
\frac{\partial^2\Gamma}{\partial\bar\sigma_{\tilde\alpha}(q)\partial
\beta_\mu^z(p)}\right.\nonumber\\
&&\left.-{\R}^{(c)^*}_{zy}(p,p')(\Gamma^{(2)}+{\cal R})^{-1}
_{\xi^y(p')\sigma_{\tilde\alpha}(q)}
\frac{\partial^2\Gamma}{\partial\bar\sigma_{\tilde\alpha}(q)\partial
\bar\gamma^z(p)}\right\}\nonumber\\
&&-\frac{1}{\alpha\bar g}\int\frac{d^dp}{(2\pi)^d}
\frac{d^dp'}{(2\pi)^d}\Biggl\{
ip_\mu {\R}^{(c)}_{zy}(p,p')(\Gamma^{(2)}+{\cal R})^{-1}_
{\bar\xi^y(p')A_\mu^z(p)}\nonumber\\
&&\left.-\bar gf_{xzw}\int\frac{d^dq}{(2\pi)^d}\bar A^x_\mu(q-p)
{\R}^{(c)}_{zy}
(p,p')(\Gamma^{(2)}+{\cal R})^{-1}_{\bar\xi^y(p')A^w_\mu(q)}
\right\}\ear
and
\bear\label{A.28}
&&{\cal A}^{(m)}_{\rm BRS}=\int\frac{d^dp}{(2\pi)^d}
\frac{d^dp'}{(2\pi)^d}\frac{d^dq}{(2\pi)^d}
\Biggl\{-{\R}^{(\psi)}_{mn}(p,p')(\Gamma^{(2)}+
{\cal R})^{-1}_{\bar\psi_n(p')\sigma_{\tilde\alpha}(q)}
\nonumber\\
&&\frac{\partial^2\Gamma}{\partial\bar\sigma_{\tilde\alpha}
(q)\partial\delta
_m^{(\psi)}(p)}-{\R}^{(\psi)*}_{mn}(p,p')
(\Gamma^{(2)}+{\cal R})^{-1}_{\psi_n
(p')\sigma_{\tilde
\alpha}(q)}\frac{\partial^2\Gamma}{\partial\bar\sigma
_{\tilde\alpha}(q)
\partial\bar\delta^{(\psi)}_m(p)}\nonumber\\
&&+{\R}^{(\varphi)}_{ab}(p,p')(\Gamma^{(2)}+
{\cal R})^{-1}_{\bar\varphi_b
(p')\sigma_{\tilde
\alpha}(q)}\frac{\partial^2\Gamma}{\partial\bar
\sigma_{\tilde\alpha}(q)
\partial\delta^{(\varphi)}_a(p)}\nonumber\\
&&+{\R}^{(\varphi)^*}_{ab}(p,p')
(\Gamma^{(2)}+{\cal R})^{-1}_{\varphi_b
(p')\sigma_{\tilde
\alpha}(q)}\frac{\partial^2\Gamma}{\partial\bar\sigma_{\tilde\alpha}(q)
\partial\bar\delta^{(\varphi)}_a(p)}\Biggr\}\ear
Here we have subtracted the ``bare'' gauge-fixing term
\be\label{A.29}
\Gamma'=\Gamma_{k_{|\gamma=0}}-\frac{1}{2\alpha}\int d^dx(D_\mu(\bar A)
(A_\mu-\bar A_\mu)^z(D_\nu(\bar A)(A_\nu-\bar A_\nu))^z\ee
and we use the index convention (cf. (\ref{A.13}))\footnote{Note
the minus sign whenever the second index is $\xi$ or $\psi$.}
\bear\label{A.29a}
&&\Gamma^{(2)}_{\bar\varphi_b
(p)\chi_{\tilde
\alpha}(q)}=\frac{\partial^2\Gamma}{\partial\varphi_b^*
(p)\partial\chi_{\tilde\alpha}(q)}\ ,\ \Gamma^{(2)}_{\varphi_b(p)
\xi_z(q)}=-\frac{\partial^2\Gamma}{\partial\varphi_b
(p)\partial\xi_z(q)}\nonumber\\
&&\Gamma^{(2)}_{\bar\psi(p)\psi(q)}=\Gamma^{(2)}_{\psi(q)\bar\psi(p)}
=\frac{\partial^2\Gamma}{\partial\psi(q)
\partial\bar\psi(p)}=-\frac{\partial^2\Gamma}{\partial\bar\psi
(p)\partial\psi(q)}\nonumber\\
&&\Gamma^{(2)}_{\psi(p)\psi(q)}
=-\frac{\partial^2\Gamma}{\partial \psi(p)\partial\psi
(q)}\ ,\ \Gamma^{(2)}_{\bar\psi(p)\bar\psi(q)}
=\frac{\partial^2\Gamma}{\partial\bar \psi(p)\partial\bar\psi
(q)}.\ear
We recover the usual identities in the limit $k\to0$ since ${\cal
R}=0$ implies a vanishing BRS-anomaly ${\cal A}_{\rm BRS}=0$.

The sources $\bar\beta$ appear only linearly in $S$ and it is
straightforward to derive the identity
\be\label{A.30}
\frac{\partial\Gamma}{\bar\beta^z_\mu}=-\frac{\partial W}{\partial
\bar\beta_\mu^z}=<\frac{\partial S}{\partial\bar\beta_\mu^z}>
=-\frac{1}{\bar g}D_\mu(A)^{zy}\xi^y-\frac{1}{\bar g}f^{zwy}<a^w_\mu
c^y>_c\ee
where the connected two-point function $<ac>_c$ is related to the
appropriate matrix element of the propagator $(\Gamma^{(2)}+
{\cal R})^{-1}$ by (\ref{A.10})
\be\label{A.30a}
<a^w_\mu c^y>_c=(\Gamma^{(2)}+{\cal R})^{-1}
_{{A^*}_\mu^w\bar\xi^y}\ee
This allows one to eliminate the explicit dependence on the
source $\bar\beta$ in favour of expressions containing two-
and three-point functions and to restrict the discussion to $\bar\beta
=0$ afterwards. Identities similar to (\ref{A.30}) for the other
sources of BRS-variations are easily derived
\bear\label{A.30b}
&&\frac{\partial\Gamma}{\partial\bar\gamma^z}=-\frac{1}{2}
f^{zyw}(\xi^y\xi^w+<c^yc^w>_c)\nonumber\\
&&\frac{\partial\Gamma}{\partial\bar\delta_a^{(\varphi)}}
=-i(T_z)_{ab}
(\xi^z\varphi_b+<c^z\varphi_b'>_c)\nonumber\\
&&\frac{\partial\Gamma}{\partial\delta_a^{(\varphi)}}=i(T_z^*)_{ab}
(\xi^z\varphi_b^*+<c^z{\varphi'}_b^*>)\ear
Finally, one has the field equation for the ghost field
\bear\label{A.31}
&&0=<\frac{\partial S}{\partial c^z}>=-\frac{\partial\Gamma}
{\partial\xi^z}-\frac{1}{\bar g}D_\mu(A)^{zy}\bar\beta^y_\mu
-f^{zwy}\xi^w\bar\gamma^y\nonumber\\
&&+i\bar\delta_m^{(\psi)}(T_z)_{mn}\psi_n
+i\delta_m^{(\psi)}(T_z^*)_{mn}\bar\psi_n+i\bar\delta_a^{(\varphi)}
(T_z)_{ab}\varphi_b
-i\delta_a^{(\varphi)}(T^*_z)_{ab}\varphi^*_b\nonumber\\
&&+D_\mu(A)^{zy}
D_\mu(\bar A)^{yw}\bar\xi^w-f^{zyv}<a^v_\mu
D_\mu(\bar A)^{yw}\bar c^w>_c\ear

For an abelian gauge theory the vanishing structure constants
$f_{zyw}$ lead to important simplifications: First one can replace
(\ref{A.30})
\be\label{A.32}
\frac{\partial\Gamma}{\partial\bar\beta_\mu}=-\frac{1}{\bar g}
\partial_\mu\xi,\quad\frac{\partial\Gamma}{\partial\bar\gamma}=0\ee
and evaluate everything for $\bar\gamma=\bar\beta_\mu=0$.
The field equations for the ghosts reduce to
\bear\label{A.33}
&&\frac{\partial\Gamma}{\partial\xi}=-i\bar\delta\varphi
+i\delta\varphi^*+\partial_\mu\partial_\mu\bar\xi\nonumber\\
&&\frac{\partial\Gamma}{\partial\bar\xi}=-\partial_\mu\partial_\mu\xi
\ear
where we have limited the matter content to a complex scalar with
$(T_z)_{ab}=-\delta_{ab}$. The ghost-dependent part of $\Gamma$
is therefore uniquely determined
\be\label{A.34}
\Gamma_{\rm gh}=\int d^dx\left\{
\partial_\mu\bar\xi\partial_\mu\xi-i\xi\bar
\delta\varphi+i\xi\delta\varphi^*\right\}\ee
It is independent of $k$ and equals the classical expression. Eq.
(\ref{A.34}) also implies that for $\bar\delta=\delta=0$ the
connected two-point functions involving $\xi$ and $\varphi',
{\varphi^*}'$ vanish
\be\label{A.35}
<c\ \varphi'>_{c_{|\bar\delta=0}}=<c\ {\varphi^*}'>_{c_{|\delta=0}}
=0\ee
and therefore
\be\label{A.36}
\frac{\partial\Gamma}{\partial\bar\delta}_{|\delta=\bar\delta=0}=
i\xi\varphi,\quad
\frac{\partial\Gamma}{\partial\delta}_{|\delta=\bar\delta=0}=
-i\xi\varphi^*\ee
We can now evaluate the Ward identiy (\ref{A.26}) for $\bar\beta
_\mu=0,\ \bar\delta=\delta=0$ where ${\cal A}_{\rm BRS}^{(g)}=0$:
\bear\label{A.37}
&&i\int\frac{d^dq}{(2\pi)^d}\left\{\frac{1}{\bar g}q_\mu\frac{\partial
\Gamma'}{\partial A_\mu(q)}+\int\frac{d^dp}{(2\pi)^d}\left[
\frac{\partial\Gamma'}{\partial\varphi(p)}\varphi(p-q)\right.\right.
\nonumber\\
&&\left.\left.-\frac{\partial\Gamma'}{\partial\varphi^*(p)}\varphi^*(p+q)
\right]\right\}\xi(q)={\cal A}^{(m)}_{\rm BRS}\ear
\bear\label{A.38}
&&{\cal A}_{\rm BRS}^{(m)}=i\int\frac{d^dp}{(2\pi)^d}
\frac{d^dp'}{(2\pi)^d}\frac{d^dq}{(2\pi)^d}
\left\{{\R}^{(\varphi)^*}(p,p')(\Gamma^{(2)}+{\cal R})^{-1}_{\varphi(p')
\bar\varphi(p-q)}\right.\nonumber\\
&&\left.-{\R}^{(\varphi)}(p,p')(\Gamma^{(2)}+{\cal R})^{-1}_
{\bar\varphi(p')\varphi(p+q)}\right\}\xi(q)\ear
Since for $\bar\delta=\delta=0$ the ghost sector decouples
completely from the $(\varphi,\varphi^*,A)$ sector we can
evaluate $(\Gamma^{(2)}+{\cal R})^{-1}$ for $\xi=\bar\xi=0$.
Furthermore the definitions (\ref{A.10}), (\ref{A.11}) imply
\be\label{A.39}
(\Gamma^{(2)}+{\cal R})^{-1}_{\varphi(p')\bar\varphi(p-q)}
=<{\varphi^*}'(p')\varphi'(p-q)>_c=(\Gamma^{(2)}+{\cal  
R})^{-1}_{\bar\varphi(p-q)\varphi(p')}
\ee
and ${\cal R}^{(\varphi)*}(p,p')={\cal R}^{(\varphi)}(p',p)$.
Since (\ref{A.37}) must hold for arbitrary $\xi(q)$
we finally obtain the modified Ward identity in a form not
involving the ghost anymore
\bear\label{A.40}
&&\frac{1}{\bar g}q_\mu\frac{\partial\Gamma'}{\partial
A_\mu(q)}+\int\frac{d^dp}{(2\pi)^d}\left[\frac{\partial
\Gamma'}{\partial\varphi(p)}\varphi(p-q)-
\frac{\partial\Gamma'}{\partial\varphi^*(p)}\varphi^*(p+q)
\right]\\
&&=\int\frac{d^dp}{(2\pi)^d}
\frac{d^dp'}{(2\pi)^d}\left[({\R}^{(\varphi)}(p,p'+q)-{\R}^{(\varphi)}
(p-q,p'))(\Gamma^{(2)}+{\cal R})^{-1}_
{\bar\varphi(p')\varphi(p)}\right]\nonumber\ear
The existence of such a form is not surprising since for our
choice of gauge fixing the ghost sector is just a free
field theory and can be omitted altogether. We observe that the
l.h.s. of eq. (\ref{A.40}) is simply a local gauge variation of
$\Gamma'$ with $\bar A$ kept fixed.
The gauge-invariant and $\bar A$-independent
part of $\Gamma'$ does therefore
not contribute. From $\Gamma_k(\varphi,A,\bar A)\equiv\Gamma_k
(\varphi,A,\bar A,\xi=\bar\xi=0)$ we may subtract a gauge-invariant
kernel $\bar\Gamma_k[\varphi,A]=\Gamma_k[\varphi,A,\bar A=A]$
and define
\be\label{A.41}
\hat\Gamma_{{\rm gf}, k}=\Gamma_k[\varphi,A,\bar A]
-\bar\Gamma_k[\varphi,A]-\Gamma_{\rm gf}[A,\bar A]\ee
We can therefore replace $\Gamma'$ by
$\hat\Gamma_{\rm gf}$ on the l.h.s.
of (\ref{A.40}). The Ward identity only constrains the
``generalized gauge-fixing term'' $\hat\Gamma_{{\rm gf},k}$
which contains
the $k$-dependent counterterms for $k>0$ and vanishes for
$k=0$. The invariance of the average action with respect to
simultaneous gauge transformations of $A$ and $\bar A$ implies
\be\label{A.42}
\frac{1}{\bar g}q_\mu\left(\frac{\partial\Gamma}{\partial A_\mu(q)}
+\frac{\partial\Gamma}{\partial\bar A_\mu(q)}\right)
+\int\frac{d^dp}{(2\pi)^d}\left[\frac{\partial\Gamma}
{\partial\varphi(p)}\varphi(p-q)-\frac{\partial\Gamma}{\partial
\varphi^*(p)}\varphi^*(p+q)\right]=0\ee
and similarly for $\Gamma'$ or $\hat\Gamma_{\rm gf}$ leading
to an alternative form of the identity (\ref{A.40})
\be\label{A.43}
q_\mu\frac{\partial\hat\Gamma_{\rm gf}}{\partial\bar A_\mu(q)}
=-\bar g\int\frac{d^dp}{(2\pi)^d}\frac{d^dp'}{(2\pi)^d}
({\R}^{(\varphi)}(p,p'+q)-
{\R}^{(\varphi)}(p-q,p'))(\Gamma^{(2)}+{\cal R})^{-1}_{\bar\varphi
(p')\varphi(p)}\ee
It is interesting to compare this equation to an identity for
the background field dependence of $\hat\Gamma_{\rm gf}$ derived
earlier \cite{Reu}
\be\label{A.44}
\frac{\partial\hat\Gamma_{\rm gf}}{\partial\bar A_\mu(q)}=
\int\frac{d^dp}{(2\pi)^d}\frac{d^dp'}{(2\pi)^d}
\frac{\partial {\R}^{(\varphi)}
(p,p')}{\partial\bar A_\mu(q)}(\Gamma^{(2)}+{\cal R}
)^{-1}_{\bar\varphi(p')\varphi(p)}\ee
One can show \cite{FreWe} that the Ward identity {\ref{A.43}) can be
derived from the more general ``background field identity''
(\ref{A.44}) by using the invariance of $\Delta^{(\varphi)}_kS$
under simultaneous gauge transformations of $\varphi, A$
and $\bar A$ which yields
\be\label{A.48}
\frac{1}{\bar g}q_\mu\frac{\partial{\cal R}^{(\varphi)}(p,p')}
{\partial\bar A_\mu(q)}={\cal R}^{(\varphi)}
(p-q,p')-{\cal R}^{(\varphi)}(p,p'+q)\ee
The background field identity (\ref{A.44}) has a simple solution
in the approximation where the $\bar A$ dependence of the
propagator $(\Gamma^{(2)}+{\cal R})^{-1}$ on the r.h.s.
is neglected
\be\label{A.49}
\hat\Gamma_{\rm gf}=<\Delta_k^{(\varphi)}S>-
\Delta_k^{(\varphi)}S[\varphi]=\int\frac{d^dp}{(2\pi)^d}
\frac{d^dp'}{(2\pi)^d}{\cal R}^{(\varphi)}(p,p')
<\varphi^*(p)\varphi(p')>_c\ee
A background field identity for non-abelian gauge theories
has also be derived \cite{Reu}, \cite{Reu2}. The precise
relation to the Slavnov-Taylor identity has not yet been
established. It is clear, however, that the background field
identity contains information beyond the Slavnov-Taylor
identity.

Let us finally turn to solutions of the Slavnov-Taylor identity
for non-abelian gauge theories. We concentrate first on a
vanishing infrared cutoff ${\cal R}_k=0$. We are interested
in solutions for vanishing sources $\beta$ for the BRS-variations.
Eq. (A.26) contains then a sum of expressions for which
the first factor can be evaluated at $\beta=0$, whereas the second
factor involves the coefficients linear in $\beta$ evaluated
at $\beta=0$.  It is straightforward to show that the ansatz
\bear\label{a}
\Gamma&=&\Gamma_{inv}[A,\psi,\varphi]+
\Gamma_{\rm gf}+\Gamma_{\rm gh}+\Gamma_s+\Delta\Gamma[\bar A]
\nonumber\\
\Gamma_{\gf}&=&\frac{1}{2\alpha}\int d^dx(D^\mu(\bar A)(A_\mu-\bar  
A_\mu))^z(D^\nu(\bar A)(A_\nu-\bar A_\nu)^z\nonumber\\
\Gamma_{\gh}&=&-\int d^dx\bar\xi^z(D^\mu(\bar A)D_\mu(A)\xi)^z
\nonumber\\
\Gamma_s&=&-\int d^dx\Bigl\{\frac{1}{\bar g}\bar\beta^z_\mu
(D^\mu(A)\xi)^z+\frac{1}{2}\bar\gamma^z f^{zyw}\xi^y\xi^w
\nonumber\\
&&-i\bar\delta_m^{(\psi)}\xi^z(T_z)_{mn}\psi_n+i\delta^{(\psi)}_m
\xi^z(T^*_z)_{mn}\bar\psi_n\nonumber\\
&&+i\bar\delta_a^{(\varphi)}\xi^z(T_z)_{ab}\varphi_b
-i\delta_a^{(\varphi)}\xi^z
(T_z^*)_{ab}\varphi^*_b\Bigr\}\ear
obeys the identity (A.26) with vanishing anomaly on the r.h.s.
Here $\Gamma_{inv}[A,\psi,\varphi]$ is an arbitrary gauge-invariant
functional which does not depend on the ghost fields, and
$\Delta\Gamma[\bar A]$ is an arbitrary gauge-invariant
functional of the background field $\bar A$. In fact, gauge
invariance of $\Gamma_{inv}$ implies the identity
\bear\label{b}
&&\int\frac{d^dp}{(2\pi)^d}\Bigl\{\frac{1}{\bar g}\frac{\partial
\Gamma_{inv}}{\partial A^z_\mu(p)}(D_\mu(A)\xi)^z(p)+i\frac
{\partial\Gamma_{inv}}{\partial\psi_m(p)}(\xi^z(T_z)_{mn}\psi_n)
(p)\nonumber\\
&&-i\frac{\partial\Gamma_{inv}}{\partial\bar\psi_m(p)}(\xi^z
(T^*_z)_{mn}\bar\psi_n)(p)\nonumber\\
&&+i\frac{\partial\Gamma_{inv}}{\partial\varphi_a(p)}(\xi^z
(T_z)_{ab}\varphi_b)(p)
-i\frac{\partial\Gamma_{inv}}{\partial\varphi_a^*(p)}(\xi^z
(T^*_z)_{ab}\varphi_b^*)(p)\Bigr\}=0\ear
This allows to replace $\Gamma'$ by $\Gamma_{\gh}$ for the
first factors in (A.26). What remains is the relation
\be\label{c}
\int\frac{d^dp}{(2\pi)^d}\left\{\frac{\partial\Gamma_{\gh}}
{\partial A^z_\mu(p)}(D^\mu(A)\xi)^z(p)-
\frac{\bar g}{2}f^{zyw}\frac{\partial\Gamma_{\gh}}{\partial
\xi^z(p)}
(\xi^y\xi^w)(p)\right\}=0\ee
which is easily verified using the Jacobi
identity for the structure
constants $f^{ztw}f^{tsy}-f^{zty}f^{tsw}
=f^{zst}f^{tyw}$. We
observe that the gauge-invariant part $\Gamma_{inv}$
remains completely unconstrained by the Slavnov-Taylor identity.
We can also verify that the ansatz (\ref{a}) obeys the field
equation for the antighost (A.24). In contrast, the source
identities (A.31), (A.33) hold only if the
pieces involving the connected two-point functions
$<ac>_c, <cc>_c,<c\varphi'>_c$ etc. all vanish.\footnote{These
two-point functions should be evaluated from the ansatz
(\ref{a}) for vanishing sources $\beta$.} This is generically
not the case for the ansatz (\ref{a}), since the cubic vertex
$\sim \bar g\bar\xi\xi A$ in $\Gamma_{\gh}$ induces a
non-trivial off-diagonal matrix element in the inverse
propagator $\Gamma_k^{(2)}$, mixing the ghost sector to
the other fields. This, in turn, is responsible for
corresponding off-diagonal elements in $(\Gamma^{(2)}+{\cal R})
^{-1}$. We observe that these off-diagonal elements vanish in
the limit of small gauge coupling $\bar g\to0$ such that
(A.31) and (A.33) are obeyed in this approximation. The
situation for the ghost field equation (A.34) is completely
analogous. We conclude that for small $\bar g\to 0$ the
whole picture becomes formally very similar to the abelian
case discussed above, with leading non-abelian structure given by
the ansatz (\ref{a}). We should emphasize that this
approximation can be expected to be valid only for the
momentum range where the running renormalized gauge coupling remains
small enough.

The ansatz (\ref{a}) is, however, not the most general solution.
We may try to obtain a more general solution by replacing in
(\ref{a}) the ghost field $\xi^z$ by a functional $\hat\xi^z
[A,\xi,\bar\xi,\psi,\varphi]$ and changing the term linear
in $\bar\gamma^z$. The functional $\hat\xi^z$ should have
ghost number one and obey the same symmetry transformation
laws as $\xi^z$ under gauge transformations acting on $A$
and $\bar A$ simultaneously. Eq. (\ref{b}) holds also with $\xi$
replaced by $\hat\xi$ such that $\Gamma_{inv}$ again drops out.
The remaining equation relates $\frac{\partial\Gamma}{\partial
\bar\gamma}$ to the functional form of $\hat\xi$
\bear\label{d}
&&\int\frac{d^dp}{(2\pi)^d}\frac{d^dq}{(2\pi)^d}(D_\mu(A)
D^\mu(\bar A)\bar\xi)^z(q)\Bigl\{\frac{\partial\hat\xi^z(q)}
{\partial\xi^w(p)}\frac{\partial\Gamma}{\partial\bar\gamma^w
(p)}\nonumber\\
&&+\frac{1}{2}f^{zyw}(\hat\xi^y\hat\xi^w)(p)(2\pi)^d\delta(q-p)
+\frac{1}{\bar  g}\frac{\partial\hat\xi^z(q)}{\partial A^w_\mu(p)}
(D^\mu(A)\hat\xi)^w(p)\nonumber\\
&&+i\frac{\partial\hat\xi^z(q)}
{\partial\psi_m(p)}(\hat\xi^y(T_y)_{mn}\psi_n)(p)-i\frac{
\partial\hat\xi^z(q)}{\partial\bar\psi_m(p)}
\left(\hat\xi^y(T_y^*)_{mn}\bar\psi_n\right)(p)\nonumber\\
&&-i\frac{\partial\hat\xi^z(q)}{\partial\varphi_a(p)}(\hat\xi
^y(T_y)_{ab}\varphi_b)(p)+i\frac{\partial\hat\xi^z(q)}
{\partial\varphi_a^*(p)}(\hat\xi^y(T^*_y)_{ab}
\varphi^*_b)(p)\Bigr\}
=0\ear
Given an arbitrary $\hat\xi$ one can always solve the equation
for $\frac{\partial\Gamma}{\partial\bar\gamma}$
provided the relation between $\hat\xi$ and $\xi$ remains
invertible. At this level we have therefore a general class of
solutions for the identity (A.26) since $\hat\xi$ remains
essentially unconstrained. The constraints arise from
the source identities (A.31), (A.33) and the ghost field
equation (A.34). (Note that the field equation for the antighost
(A.24) is obeyed for arbitrary $\hat\xi$.) In fact, the difference
between $\hat\xi$ and $\xi$ is related to the connected two-point
functions involving the ghost field
\bear\label{e}
&&(D_\mu\hat\xi)^z-(D_\mu\xi)^z=\bar gf^{zwy}<a^w_\mu c^y>_c
\nonumber\\
&&(T_z)_{ab}((\hat\xi^z-\xi^z)\varphi_b-<c^z\varphi_b'>_c)=0\ear
Similar relations must hold for $\varphi^*,\psi$
and $\bar\psi$ and the equation for $\frac{\partial\Gamma}
{\partial\bar\gamma}$ must be compatible with the solution
of (\ref{d}). We conclude that the
generalized ansatz with $\xi$ replaced by $\hat\xi$ can only be
used in the approximation where this system
of equations for $\hat\xi$ is
self-consistent. Beyond this approximation the gauge invariance
of the sector with ghost number zero $(\Gamma_{inv})$ can
probably not be maintained.

In the presence of a nonvanishing infrared cutoff ${\cal R}_k$
an additional piece $\Gamma_{ct}$ containing counterterms
should be added. It vanishes in the limit $k\to0$. In
the perturbative regime $\Gamma_{ct}$ contains a gluon mass
term $\sim k^2$ (see sect. 4). There is no reason why $\Gamma_{ct}$
should be gauge-invariant under transformations which leave
the background field $\bar A$ fixed. We may define $\Gamma_{ct}$
by the requirement that $\Gamma-\Gamma_{ct}$ obeys the anomaly-free
Slavnov-Taylor identity. The anomaly $A^{(g)}_{\rm BRS}+A^{(m)}
_{\rm BRS}$ on the r.h.s. of eq. (A.26) determines then the
form of $\Gamma_{ct}$. In a lowest order approximation we may
treat $\Gamma_{ct}$ and the anomaly as a small quantity and linearize
eq. (A.26) in $\Gamma_{ct}$.

For the computation of the gluon propagator in the present
paper we make only a crude approximation for the three- and
four-gluon vertices appearing on the r.h.s. of the flow equation.
They correspond to the ansatz (\ref{a}) with $\Gamma_{inv}$
containing only a piece $\sim F_{\mu\nu}F^{\mu\nu}$. An
improved treatment of these vertices could generalize $\Gamma_
{inv}$ for a nontrivial momentum dependence of the propagator
\be\label{f}
\Gamma_{inv}=\frac{1}{4}\int dx F_{\mu\nu}K(-D^2(A))F^{\mu\nu}\ee
with $K(x)=(G_A(x)-G_A(0)))/x$.

\section*{Figure Captions}

\begin{description}
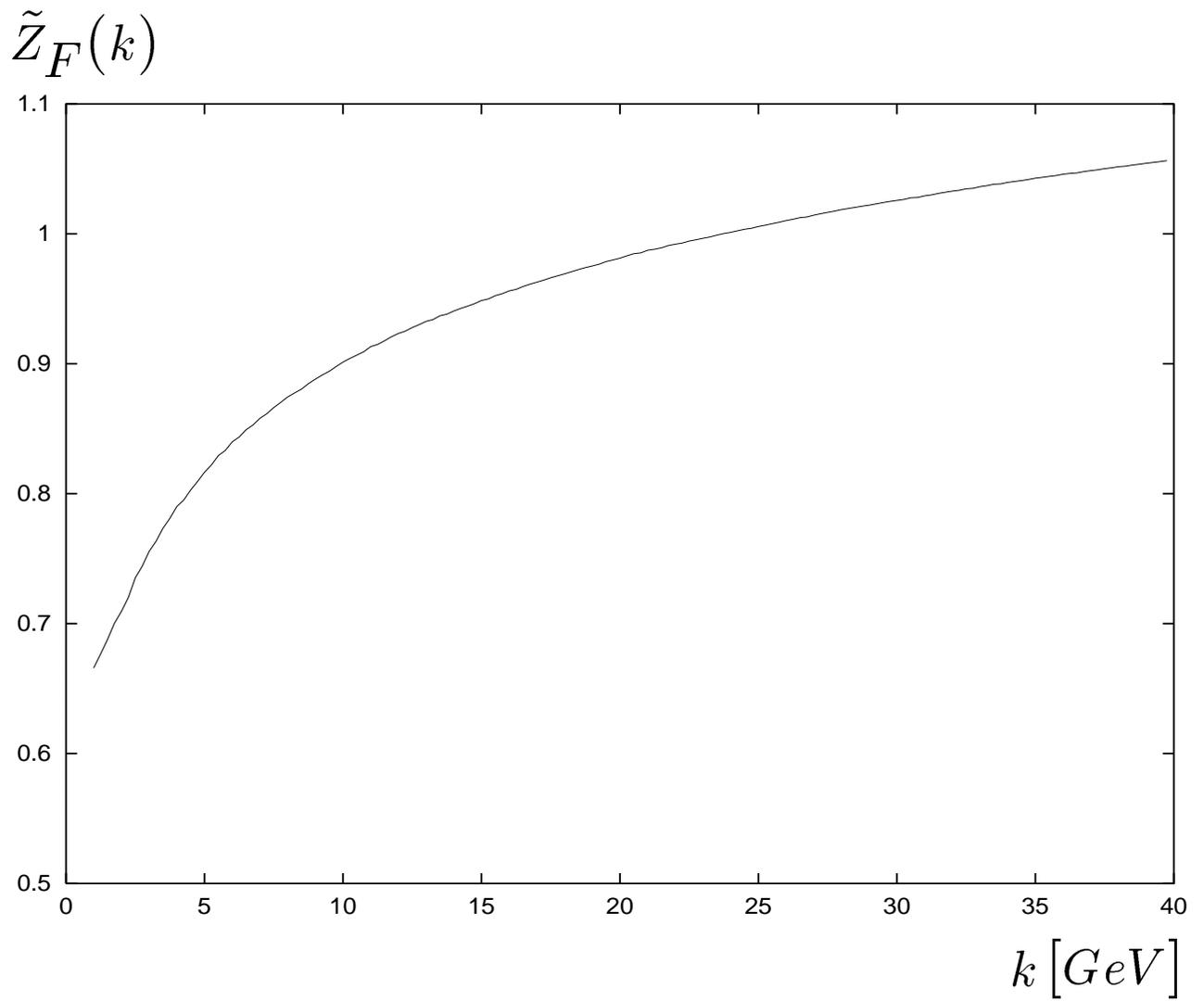
\item{Fig. 1:} Wave function renormalization $\tilde Z_F$ as
a function of the average scale $k$.
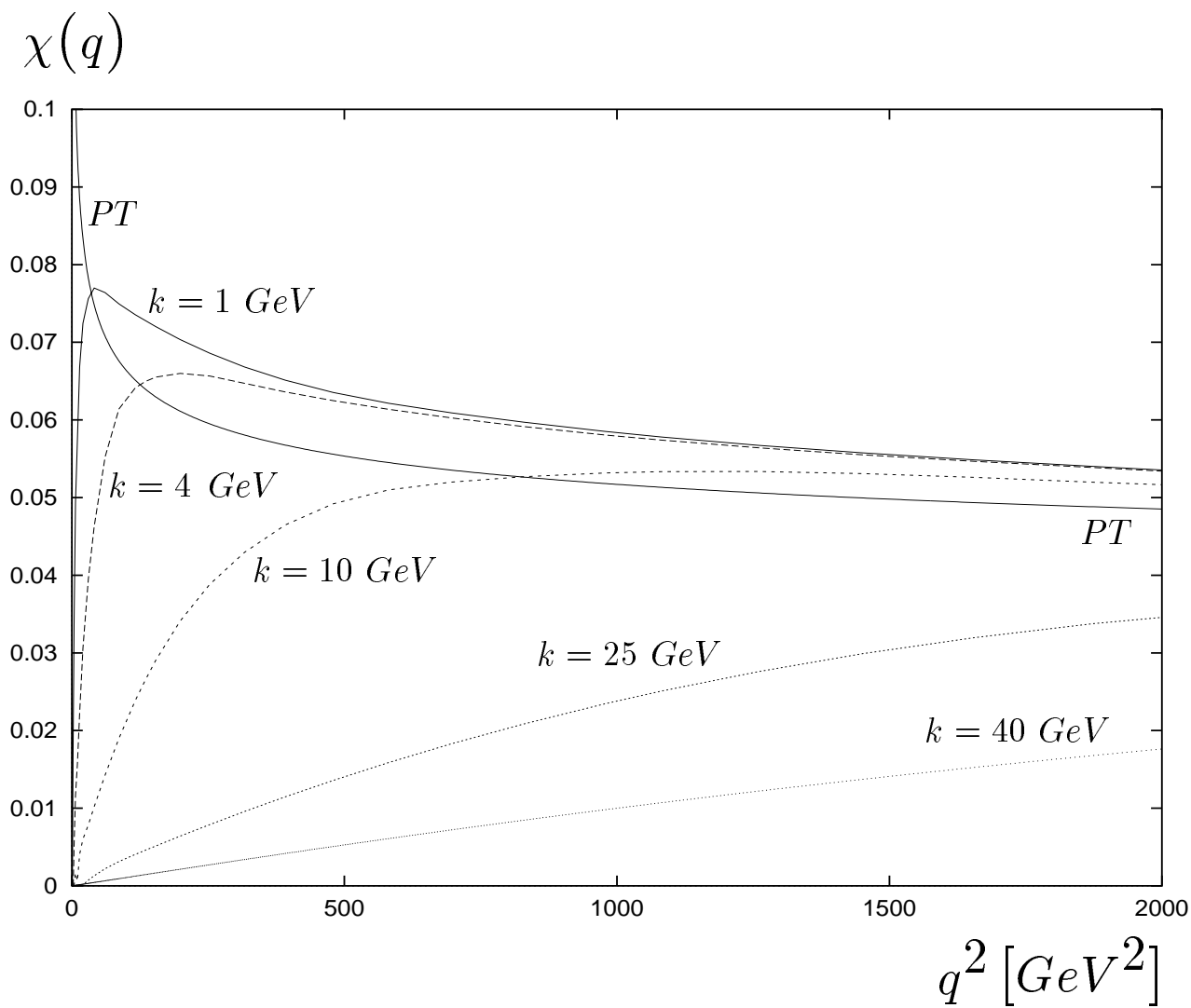
\item{Fig. 2:} Momentum-dependent anomalous dimension $\chi(q)$
in comparison with perturbation theory for various values
of $k\geq 1$ GeV.
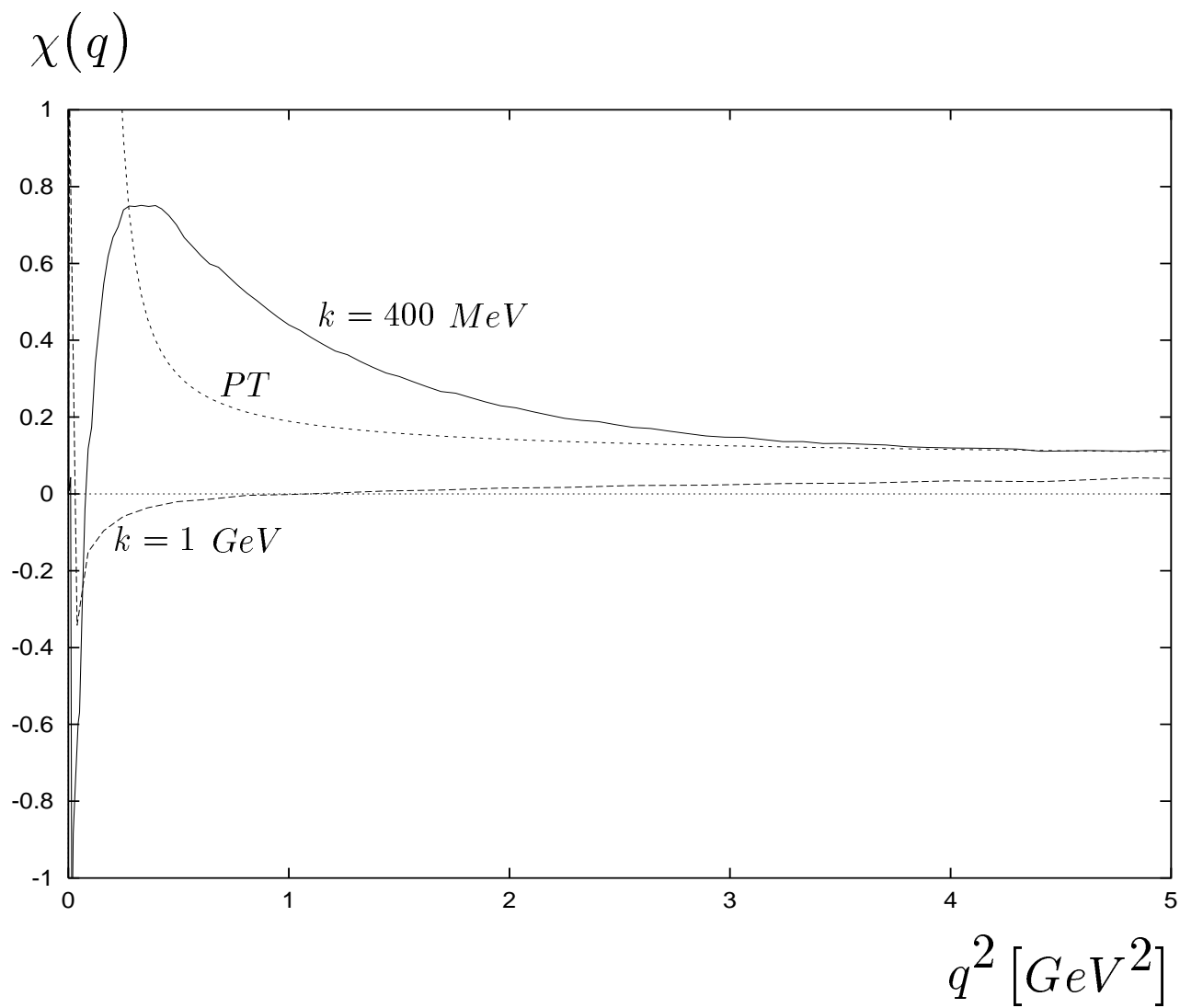
\item{Fig. 3:} The same as fig. 2, for $k\leq1$ GeV.
\end{description}

\end{document}